\documentclass[twocolumn,showpacs,preprintnumbers,amsmath,amssymb,superscriptaddress]{emulateapj}

\usepackage{graphicx}

\usepackage{aas_macros}
\usepackage{amsmath}
\usepackage{cases}

\usepackage{color}

\begin{document}


\title{Thermodynamics and Charging of Interstellar Iron
  Nanoparticles}%

\author{Brandon S. Hensley}
\email{brandon.s.hensley@jpl.nasa.gov}
\affiliation{Jet Propulsion Laboratory, California Institute of Technology, 4800
Oak Grove Drive, Pasadena, CA 91109, USA}

\author{B. T. Draine}
\affiliation{Department of Astrophysical Sciences,  Princeton
  University, Princeton, NJ 08544, USA}

\date{\today}

\begin{abstract}
Interstellar iron in the form of metallic iron nanoparticles may
constitute a component of the interstellar dust. We compute the
stability of iron nanoparticles to sublimation in the interstellar
radiation field, finding that iron clusters can persist down to a
radius of $\simeq 4.5\,$\AA, and perhaps smaller. We employ laboratory data on small iron
clusters to compute the photoelectric yields as a function of grain
size and the resulting grain charge distribution in various
interstellar environments, finding that iron
nanoparticles can acquire negative charges particularly in regions
with high gas temperatures and ionization fractions. If $\gtrsim 10\%$
of the interstellar iron is in the form of ultrasmall iron clusters,
the photoelectric heating rate from dust may be increased by up to tens of
percent relative to dust models with only carbonaceous and silicate
grains.
\end{abstract}

\section{Introduction}
More than 95\% of interstellar iron is depleted from the gas phase
\citep{Jenkins_2009}, suggesting that iron is the largest elemental
contributor to the interstellar dust mass after oxygen and carbon
\citep[e.g.][Table 23.1]{Draine_2011}. Silicate grains provide a
possible reservoir for the iron in the form of interstellar pyroxene
(Mg$_x$Fe$_{1-x}$SiO$_3$) or olivine (Mg$_{2x}$Fe$_{2-2x}$SiO$_4$)
analogues. However, the shape and strength of the 9.7\,$\mu$m silicate
feature in extinction suggest that the silicate
material is magnesium-rich rather than iron-rich
\citep{Poteet+Whittet+Draine_2015} and therefore that a substantial
fraction ($\sim$70\%) of the interstellar iron is in other forms such as iron oxides
\citep[e.g. Fe$_3$O$_4$,][]{Cox_1990, Jones_1990, Henning+etal_1995,
  Draine+Hensley_2013}, iron sulfides \citep[e.g. FeS,][]{Bradley_1994,
  Kohler+Ysard+Jones_2014}, or metallic iron \citep{Schalen_1965,
  Chlewicki+Laureijs_1988}.

Metallic Fe inclusions have been found in interplanetary dust particles
\citep{Bradley_1994}, lunar soil samples \citep{Keller+McKay_1997},
and, most recently, in putative interstellar grains collected in the
Solar System \citep{Westphal+etal_2014, Altobelli+etal_2016}. Thus, metallic iron is a
particularly compelling candidate material whether in the form of
inclusions in larger grains or, as we focus on in this work,
free-flying nanoparticles.

Because metallic Fe has no infrared or UV resonances, IR and UV
spectroscopy does not place limits on the fraction of interstellar Fe
that is metallic. \citet{Paerels+etal_2001}
concluded that X-ray spectroscopy near the Fe L$_2$ and L$_3$ edges
agreed better with metallic Fe than with Fe oxides, but
\citet{Valencic+Smith_2013} conclude that the fits are not
particularly good, and other Fe compounds can fit the L$_{2,3}$ edges
nearly as well. A large fraction of interstellar Fe could be in metallic Fe
nanoparticles without having noticeably affected the interstellar
extinction curve \citep[see Figure 13
of][]{Draine+Hensley_2013}.

Nevertheless, metallic iron can greatly impact observed dust
properties. If metallic Fe is present in the form of inclusions, it
could contribute to the alignment
of interstellar dust grains: alignment by magnetic dissipation in 
spinning grains, as proposed by \citet{Davis+Greenstein_1951}, 
could be faster if ferromagnetic materials are present
\citep{Jones+Spitzer_1967, Duley_1978, Hoang+Lazarian_2016b}. Both
free-flying and included Fe nanoparticles emit thermal magnetic dipole
radiation at submillimeter and millimeter wavelengths
\citep{Draine+Lazarian_1999, Draine+Hensley_2013}, and such grains have
been invoked to explained the observed excess submillimeter-millimeter
emission in the Small Magellanic Cloud
\citep{Draine+Hensley_2012}. Magnetic dipole emission from aligned
metallic iron grains or from iron inclusions in aligned grains is
polarized orthogonally to the electric dipole emission from
non-magnetic grains. Thus, the presence of metallic Fe grains could
dramatically affect the polarized Galactic dust spectral energy
distribution (SED) by effectively ``diluting'' the polarized electric
dipole emission. The {\it Planck} satellite has observed that the
polarization fraction of the dust emission declines with increasing
wavelength \citep{Planck_Int_XXII}, consistent with predictions from
models of magnetic dust
\citep{Draine+Hensley_2013}. 

In addition to affecting far-infrared dust emission and polarization,
stochastically-heated free-flying Fe nanoparticles may contribute
significantly to the infrared dust emission near
20\,$\mu$m. Photoelectric heating is dominated by the smallest grains
\citep[e.g.][]{Bakes+Tielens_1994, Weingartner+Draine_2001b}, and so a
population of iron nanoparticles may contribute to interstellar heating. Finally,
Fe nanoparticles may be responsible for part or
all of the so-called anomalous microwave emission (AME) peaking
near $\sim$30\,GHz either through thermal magnetic dipole emission
\citep{Draine+Lazarian_1999, Draine+Hensley_2013} or rotational emission
\citep{Hoang+Lazarian_2016}. In a separate study
\citep[][in prep.]{Hensley+Draine_2016b}, we examine the mid-infrared
emission from Fe nanoparticles undergoing stochastic heating, and from
this we obtain upper limits on the fraction of the observed AME that
could be due to rotational emission from Fe nanoparticles.

Given the potential importance of Fe nanoparticles as a component of
interstellar dust, in this work we investigate the thermodynamic and
photoelectric properties of these grains. In Section~\ref{sec:thermo},
we discuss the heating and sublimation rate of Fe nanoparticles and compute
a minimum grain size that is stable to sublimation; in
Section~\ref{sec:charge}, we discuss the collisional charging and
photoelectric emission from these grains and compute the resulting
grain charge distribution as a function of grain size; in
Section~\ref{sec:photo_e}, we quantify the contribution of these grains to the
photoelectric heating in various interstellar environments; finally, we summarize our
principal conclusions in Section~\ref{sec:conclusion}. 

\section{Grain Heating and Sublimation}
\label{sec:thermo}

\subsection{Destruction by Desorption}

\subsubsection{Thermal Energy vs.\ $T$}
\begin{figure}[ht]
\begin{center}
\includegraphics[angle=0,width=8.0cm,
                 clip=true,trim=0.5cm 5.0cm 0.5cm 2.5cm]
                {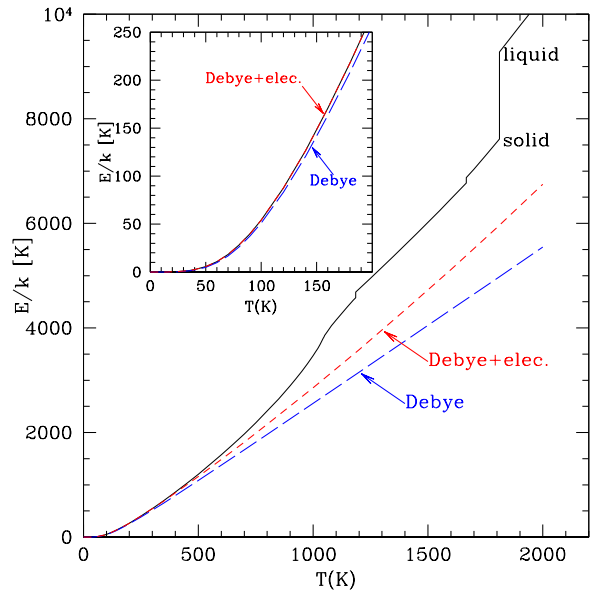}
\caption{\label{fig:fheat}\footnotesize
         $E/k$, where $E$ is the heat of formation per atom 
         of bulk Fe, relative
         to $T=0\,{\rm K}$, from \citet{Desai_1986}.
         The discontinuities at $T=1185\,{\rm K}$ and $1667\,{\rm K}$ are the
         $\alpha-\gamma$ and $\gamma-\delta$ phase transitions,
         and the larger discontinuity at $T=1811\,{\rm K}$ is the solid-liquid
         phase transition.
         The dashed red line is the Debye model plus electronic
         specific heat (Equation~\ref{eq:Emodel}).}
\end{center}
\end{figure}

The experimentally-measured heat content of bulk Fe as measured by
\citet{Desai_1986} is shown in Figure~\ref{fig:fheat}.
At low temperature ($T< 800\,{\rm K}$) the internal energy per atom
of bulk Fe is
well-approximated by a Debye model  plus an electronic heat capacity:

\begin{equation} \label{eq:Emodel}
\frac{E_{\infty}(T)}{k} = 
9\Theta_{\rm D}\int_0^1 \frac{x^3 dx}{\exp(x\Theta_{\rm D}/T)-1} +
 \frac{\gamma}{2} T^2
~~~,
\end{equation}
where
the Debye temperature
$\Theta_{\rm D}=415\,{\rm K}$,
and $\gamma=6\times10^{-4}\,{\rm K}^{-1}$.
Equation~\ref{eq:Emodel} is plotted in Figure \ref{fig:fheat}.
It fits the experimental data well for $T<300\,{\rm K}$, but at higher
temperatures the experimental heat capacity is larger than given
by Equation~\ref{eq:Emodel}.
We use the laboratory data in Figure \ref{fig:fheat} to relate
the internal energy $E$ to the temperature $T$.

\subsubsection{Vapor Pressure}

\begin{figure}[ht]
\begin{center}
\includegraphics[angle=0,width=8.0cm,
                 clip=true,trim=0.5cm 5.0cm 0.5cm 2.5cm]
{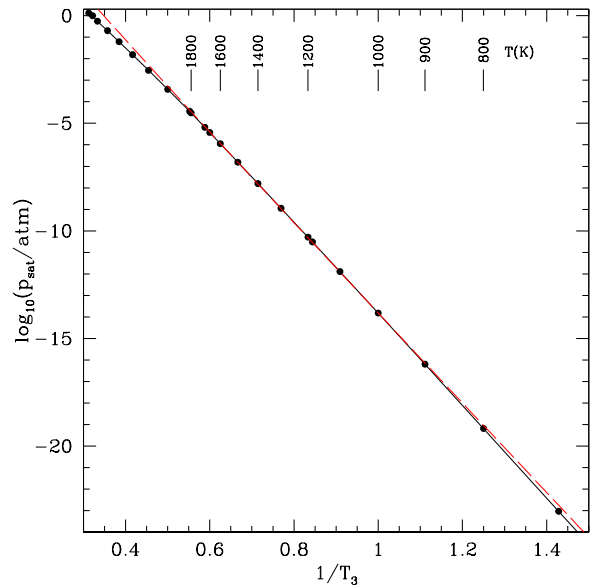}
\caption{\label{fig:svp}\footnotesize
         Solid curve: experimental saturation vapor pressure of Fe
         \citep{Desai_1986}.
         Dashed: Equation~\ref{eq:svp fit}.
         }
\end{center}
\end{figure}

The saturation vapor pressure of bulk Fe is approximated by
(see Figure \ref{fig:svp}):
 
\begin{equation} \label{eq:svp fit}
p_{\rm sat}(T)=4.80\times10^{12}T_3 \exp(-47.18/T_3)
\,{\rm erg}\,{\rm cm}^{-3}
\end{equation}
where $T_3\equiv T/10^3\,{\rm K}$,
so that the saturation concentration is

\begin{equation}
c_1^{\rm sat} = 
\frac{p_{\rm sat}}{kT} = 3.48\times10^{25}\exp(-47.18/T_3)
\,{\rm cm}^{-3} 
~~~.
\end{equation}

The saturation vapor pressure for monomers in equilibrium with clusters of 
Fe$_n$ atoms will
be greater than the bulk value, because the binding energy per atom
is lowered by surface curvature (i.e., surface free energy).
In LTE, the rate per surface area 
of sublimation of monomers from a cluster of $n$ atoms is
estimated to be
\citep{Guhathakurta+Draine_1989}

\begin{equation}
\label{eq:evaporation_rate}
R_{n}(T) \approx \alpha_{n-1} c_1^{\rm sat}
\left(\frac{kT}{2\pi m}\right)^{1/2} 
\exp\left(\frac{\theta_\infty/T}{(n-1)^{1/3}}\right)
~~~,
\end{equation}
where $\alpha_{n-1}$ is the sticking coefficient for Fe atoms
colliding with Fe$_{n-1}$, and

\begin{equation}
\theta_\infty = \left(\frac{32\pi}{3}\right)^{1/3}
\left(\frac{m}{\rho}\right)^{2/3} \frac{\sigma}{k}
~~~,
\end{equation}
where $\sigma$ is the surface free energy.
Experimental data on molten Fe is approximated by
$\sigma\approx [1918 - 0.43(T/\,{\rm K}-1811)]\,{\rm erg}\,{\rm cm}^{-2}$
\citep{Kasama+etal_1983}.
The surface free energy of solid bcc Fe at low temperatures
is estimated to be 
$\sigma\approx 2410\,{\rm erg}\,{\rm cm}^{-2}$ \citep{Tyson+Miller_1977}.
For the temperatures $T\approx1200\,{\rm K}$ of interest for Fe sublimation,
we will take $\sigma\approx2200\,{\rm erg}\,{\rm cm}^{-2}$,
corresponding to
\begin{equation}
\theta_\infty \approx 2.67\times10^4\,{\rm K}
~~~.
\end{equation}
A cluster Fe$_n$ has surface area
$A_{n}\approx 2.51\times10^{-15}n^{2/3}\,{\rm cm}^2$.
If we take $\alpha_n\approx 0.5$, then in LTE the rate coefficient for
${\rm Fe}_{n}\rightarrow {\rm Fe}_{n-1} + {\rm Fe}$ is

\begin{align}
k_{n\rightarrow n-1}^{\rm LTE} \approx\ &6.71\times10^{14} n^{2/3}
                                 T_3^{1/2} \nonumber \\
&\times\,\exp\left[-\frac{47.18}{T_3} +
  \frac{26.7}{T_3}(n-1)^{-1/3}\right]\,{\rm s}^{-1}
~~~.
\end{align}
The effective energy $B$ required to liberate one Fe atom from
a cluster Fe$_{n}$ is

\begin{align}
B &\approx k \left[47180\,{\rm K} - \frac{26700\,{\rm
    K}}{(n-1)^{1/3}}\right] \nonumber \\
\label{eq:subl}
&= \left[4.07 - 2.30 (n-1)^{-1/3}\right]\,{\rm eV}
~~~.
\end{align}
A cluster containing a fixed energy $E$ is {\it not} in LTE with a heat bath.
Fluctuations must concentrate an energy $>B$ into a single
vibrational mode to break the bond holding one of the surface
Fe atoms to the rest of the cluster.
The sublimation rate is suppressed below the LTE rate by
a factor that can be estimated using ``quantum RRK'' theory
for unimolecular reaction rates \citep{Robinson+Holbrook_1972}.
The suppression factor is estimated to be
\citep{Guhathakurta+Draine_1989}

\begin{equation} \label{eq:S_N}
S_{n} =
\left(1+\frac{1}{\gamma}\right)^b
\frac{\Gamma(\gamma f+1)}{\Gamma(\gamma f+f)}
\frac{\Gamma(\gamma f + f - b)}{\Gamma(\gamma f + 1 - b)}
~~~,
\end{equation}
where $f=3(n-2)$ is the number of vibrational degrees of freedom
of the Fe$_{n}$ cluster,

\begin{equation}
\gamma \equiv \frac{E/f}{\hbar\omega_0}
\end{equation}
is internal energy per mode divided by $\hbar\omega_0$,
and

\begin{equation}
b \equiv \frac{B}{\hbar\omega_0}
\end{equation}
is the binding energy divided by the vibrational quantum.
We take $\hbar\omega_0=(3/4)k\Theta_{\rm D}$, the mean
mode energy for a 3-dimensional Debye spectrum.
Thus, consider a nanoparticle Fe$_{n}$ undergoing stochastic
heating and cooling, with $P_E$ being the probability
of being found with internal energy $E$.
We take the sublimation rate
to be

\begin{equation} \label{eq:dNdt}
\left(\frac{dn}{dt}\right)_E =
 - \sum_E P_E S_n(E)\times k_{n\rightarrow n-1}^{\rm LTE}(T_E)
~~~,
\end{equation}
where $S_n(E)$ is given by Equation~\ref{eq:S_N},
the ``effective temperature'' 
$T_E$ is the temperature where bulk Fe
has the same mean energy per degree of
freedom as the cluster with energy $E$:

\begin{equation}
E_\infty(T_E) = \frac{E}{n-2}
~~~,
\end{equation}
and the dimensionless quantities $b$, $f$, and $\gamma$ are
\begin{align}
b &= 144.6 - 71.1 (n-1)^{-1/3}
\\
f &= 3(n-2)
\\
\gamma &= \frac{E/f}{(3/4)k\Theta_{\rm D}}
~~~.
\end{align}

\subsection{Distribution Function for Internal Energy}
The internal energy of a cluster varies with time, with upward spikes
following photon absorption, and downward transitions as the result
of thermal emission (i.e., spontaneous decays).  We divide the range
of likely internal energies into a finite number (499) of energy bins.
The time-averaged
probability $P_j$ of finding a Fe$_n$ cluster in energy bin $E_j$
is obtained by finding the steady state solution $P_i$ satisfying

\begin{equation}
0 = \sum_{j\neq i} T_{ji}P_i   \quad \quad 1 =\sum_j P_j
\end{equation}
where the transition matrix $T_{ji}$ is the probability per unit time
of a cluster in state $i$ making a transition $i\rightarrow j$.
Transition matrix elements $T_{ji}$ for downward transitions were
calculated with infrared emission treated as discrete transitions
\citep{Draine+Li_2001}. Upward $T_{ji}$ were calculated using
photoabsorption cross sections derived from the dielectric function
for metallic Fe from \citet[][Appendix B]{Draine+Hensley_2013}.

\begin{figure*}[ht]
\begin{center}
\includegraphics[angle=0,width=7.0cm,
                 clip=true,trim=0.5cm 5.0cm 0.5cm 2.5cm]
                {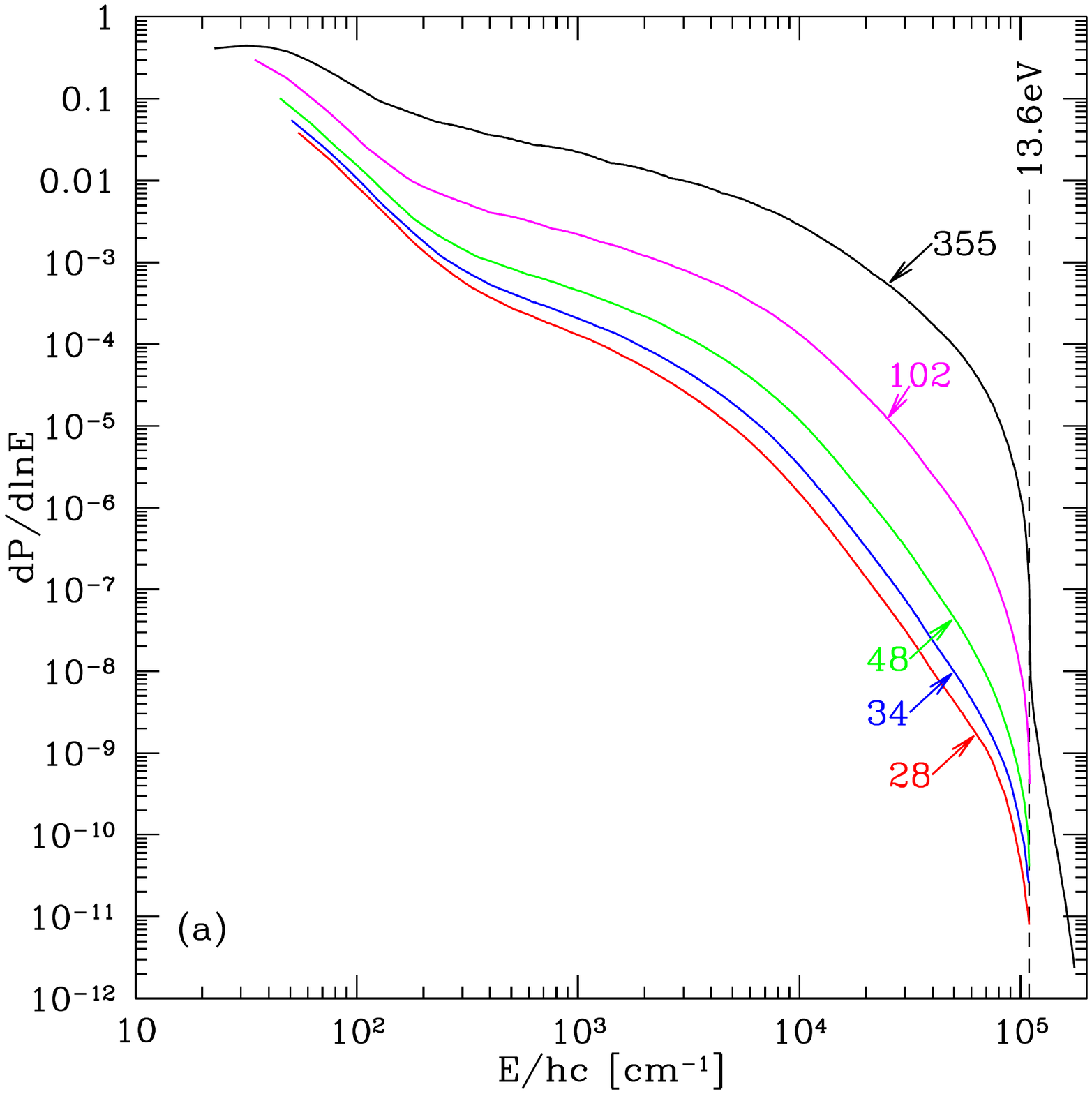}
\includegraphics[angle=0,width=7.0cm,
                 clip=true,trim=0.5cm 5.0cm 0.5cm 2.5cm]
                {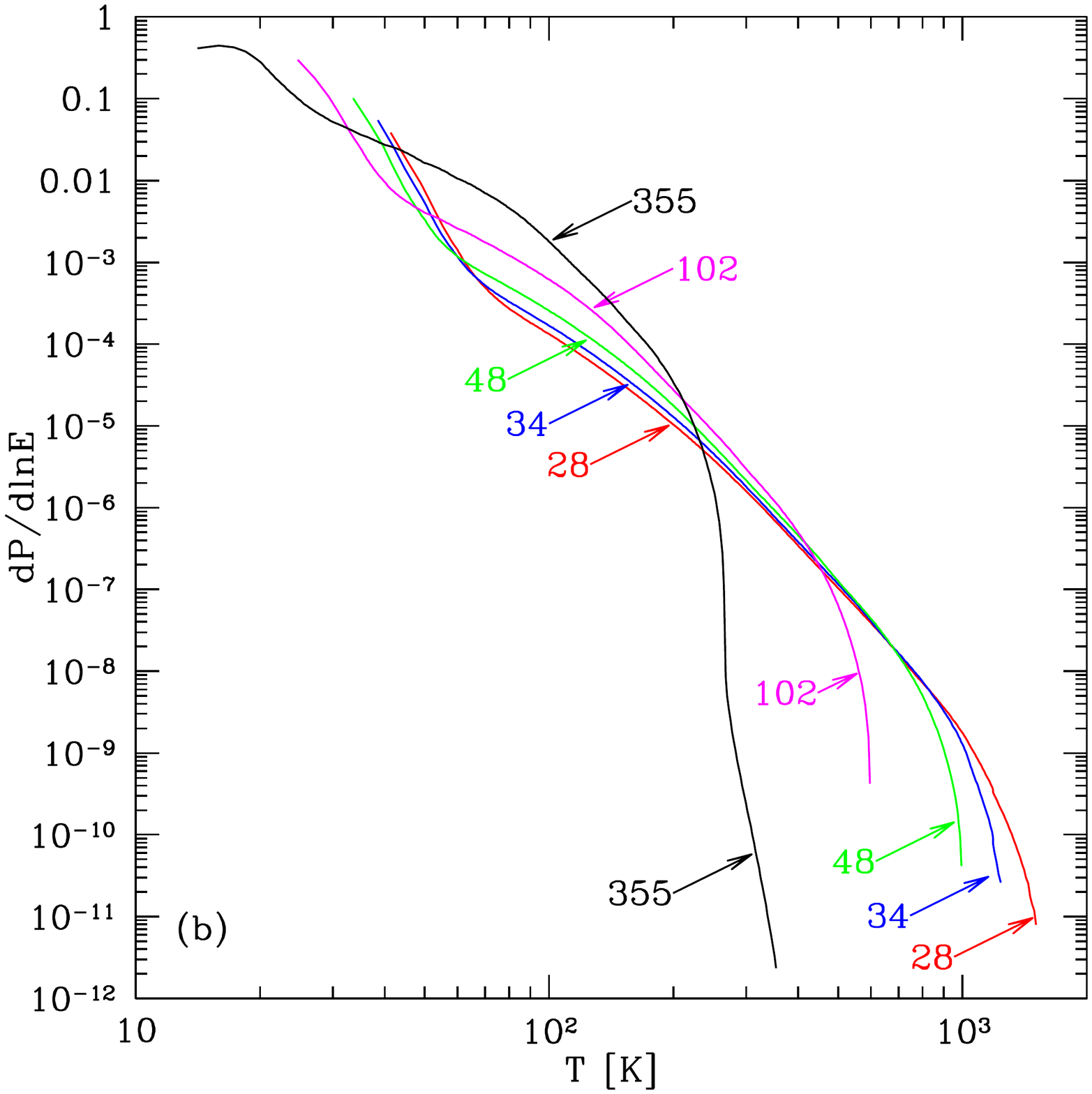}
\caption{\label{fig:dpde}\footnotesize
    Energy distribution function for Fe$_n$ nanoparticles
    vs.\ energy (left) and temperature
    (right).
    Curves are labelled by number $n$ of Fe atoms in the nanoparticle.
    }
\end{center}
\end{figure*}

Examples of energy distribution functions are shown in Figure~\ref{fig:dpde}
for Fe nanoparticles heated by the \citet{Mathis+Mezger+Panagia_1983}
interstellar radiation field.
The results shown in Figure~\ref{fig:dpde} were obtained assuming that
the only cooling process is thermal emission of photons.

The distribution functions in Figure~\ref{fig:dpde} all show a dramatic drop at
$E/hc=1.097\times10^5\,{\rm cm}^{-1}$, corresponding to $E=13.60\,{\rm eV}$, because
the starlight spectrum is assumed to be cut off above this energy.
Following a photon absorption, a cluster cools very rapidly and is likely
to be in the ground vibrational state (or very close to it) when the
next photon absorption event occurs.
Even for a cluster with as many as 355 atoms, the probability of having
an appreciable internal energy $E>1\,{\rm eV}$ is $\sim 10^{-3}$, but this is enough
for the internal distribution $P_E$ to begin to
extend noticeably above the $13.6\,{\rm eV}$ cutoff (see Figure~\ref{fig:dpde}a).
To illustrate what grain temperatures are reached during the ``thermal
spikes'' due to photon heating events, Figure~\ref{fig:dpde}b
shows $dP/d\ln E$ vs.\ the grain ``temperature'' $T_E$.

Grains with $n\lesssim50$ reach peak temperatures $T>10^3\,{\rm
  K}$. While the melting temperature of bulk Fe is $T=1811\,{\rm K}$,
melting point depression due to small particle effects can lower the
melting point below 900\,K \citep{Duan+etal_2007}. Thus, hot Fe$_n$
clusters may be liquid at their highest temperatures.\footnote{Of
  course, the notion of ``melting'' is itself 
not well-defined for small atomic clusters.} The energy required to
effect these phase transitions would, for the same heat capacities,
slightly decrease the peak temperature attained by these grains;
however, whether the grain is solid or liquid does not affect our
estimate of the evaporation rate from the grain's surface
(Equation~\ref{eq:evaporation_rate}). Note that the saturation vapor
pressure for bulk Fe is continuous across the solid-liquid phase
transition (see Figure~\ref{fig:svp}).

In the diffuse cold neutral medium (CNM), only about 1\% of the Fe is in the gas phase:
$n({\rm Fe^+})\approx1\times10^{-5}\,{\rm cm}^{-3}$, and thermal Fe$^+$ ions
would collide with a neutral Fe$_{30}$ cluster at a rate \citep{Draine+Sutin_1987}

\begin{equation}
    \left(\frac{dn}{dt}\right)_{\rm coll} \approx
    3\times10^{-9}\,{\rm cm}^3\,{\rm s}^{-1} n({\rm Fe}^+) \approx
    3\times10^{-14}\,{\rm s}^{-1}
\end{equation}
corresponding to a very long mass doubling time
$\sim$$3\times10^7$\,yr.  Growth by accretion from the gas is very
slow.  Replenishment of the mass in very small clusters is instead
probably dominated by fragmentation of larger grains.

\begin{figure}[ht]
\begin{center}
\includegraphics[angle=0,width=8.0cm,
                 clip=true,trim=0.5cm 5.0cm 0.5cm 2.5cm]
                {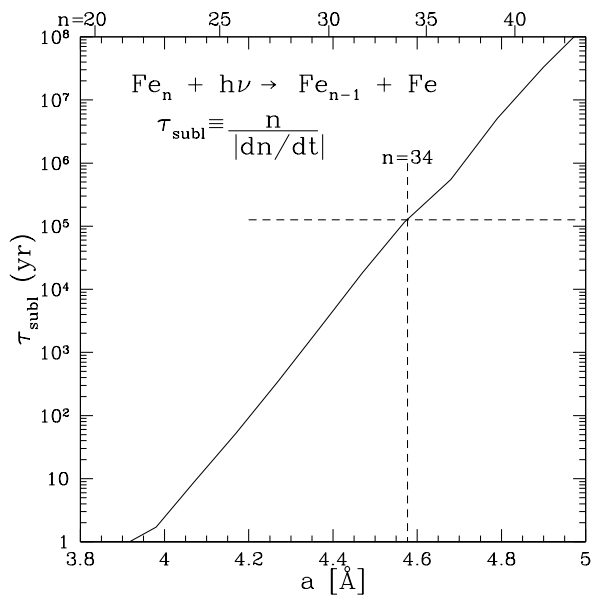}
\caption{\label{fig:tevap}\footnotesize
    Lifetime against thermal sublimation for Fe nanoparticles
    heated by the interstellar radiation field.
    }
\end{center}
\end{figure}

For each grain size, we have calculated the thermal sublimation rate
using Equation~\ref{eq:dNdt}.
The resulting timescales for thermal sublimation are shown,
as a function of grain size, in Figure~\ref{fig:tevap}.
We see from Figure~\ref{fig:tevap} that
a Fe$_{34}$ nanoparticle has
a lifetime against destruction by thermal sublimation
$\tau_{\rm subl}\approx10^5\,{\rm yr}$, and this lifetime drops to only
$\sim10^4\,{\rm yr}$ for Fe$_{30}$.
Thus, unless important cooling processes have been neglected,
we expect that Fe nanoparticles in the ISM will contain 
$n\geq34$ atoms, i.e., will have radii $a>4.5\,{\rm \AA}$ (see
Equation~\ref{eq:n_atoms}).

From Figure~\ref{fig:dpde} we see that clusters with $n=34$ and $n=48$ have
very similar temperature distribution functions for $T<1000\,{\rm K}$.
The $n=34$ cluster has a much shorter lifetime, however, because its
temperature distribution function extends up to $\sim1300\,{\rm K}$.
From this we see that the sublimation that limits the sizes of
Fe nanoparticles takes place at temperatures $T\approx
1200\pm200\,{\rm K}$. Note that our empirical fit to the saturation
vapor pressure closely reproduces experimental data (for bulk Fe) over
this range.

\subsection{Other Cooling Processes}

\subsubsection{Sublimation Cooling?}

The above calculation of $P_E$ assumed that the only cooling process
for a cluster is thermal emission.
Of course, sublimation of an Fe atom
(Fe$_n$$\rightarrow$Fe$_{n-1}$+Fe, Equation~\ref{eq:subl})
involves breaking a bond, and thus leaves the Fe$_{n-1}$ cluster in a
cooler state from which a second sublimation event will be
extremely unlikely.
We have seen above that sublimation is only important for
$T>10^3\,{\rm K}$.
At $T>10^3\,{\rm K}$, the radiative cooling time for a Fe nanoparticle is
$t_{\rm cool}\sim10^{-2}\,{\rm s}$.
The total rate of starlight absorption events for a cluster Fe$_n$ is 

\begin{equation}
R_\star \approx 5\times10^{-9}U n \,{\rm s}^{-1}
\end{equation}
where the dimensionless factor
$U$ is the intensity of the starlight relative to the
\citet{Mathis+Mezger+Panagia_1983} estimate for the local interstellar
radiation field.
The rate of sublimation is

\begin{equation}
R_{\rm sub} = 3\times10^{-13} 
n \left(\frac{\tau_{\rm subl}}{10^5\,{\rm yr}}\right)^{-1}\,{\rm
  s}^{-1}
~~~.
\end{equation}
The number of sublimations per photon absorption

\begin{equation}
\frac{R_{\rm subl}}{R_\star} 
\approx 6\times10^{-5}U^{-1}\left(\frac{\tau_{\rm subl}}{10^5\,{\rm yr}}\right)^{-1}
\ll 1
~~~.
\end{equation}
Only a very small fraction of the photon absorption events result in
a sublimation, even for grains that are near our estimated
minimum cutoff (sublimation lifetime $\tau_{\rm subl}\approx10^5\,{\rm
  yr}$).  
Neglecting sublimation cooling, this means that only
a fraction $(R_{\rm subl}/R_\star)$ of the sublimations would occur
in thermal spikes with two independent sublimations.
Thus our neglect of sublimation cooling has resulted in only a slight
overestimate of the sublimation rate.

\subsubsection{Thermionic Emission Cooling?}

A hot grain can also lose energy by emitting electrons -- the process
known as ``thermionic emission''.

A neutral Fe$_n$ nanoparticle can emit an electron

\begin{equation}
{\rm Fe}_n + \Delta E \rightarrow {\rm Fe}_n^+ + e^-
\end{equation}
if the energy $\Delta E$ exceeds the ionization potential IP.
\citet{Yang+Knickelbein_1990} have measured ionization potentials for
Fe$_n$ with $5\leq n \leq 90$ (see Figure~\ref{fig:ip}), and the experimental value of the
ionization potential for $n=34$ is $5.06\,{\rm eV}$.
Thus the cost of ejecting an electron from Fe$_{34}$ is greater than the
binding energy $B\approx 3.4\,{\rm eV}$ of a Fe atom (see
Equation~\ref{eq:subl}). The ionization potential for
Fe$_n^+$ cations is even greater, and thermionic emission is even less
likely. For a Fermi velocity of order
1000\,km\,s$^{-1}$, the trial frequency would be of order
4$\times10^{16}$\,Hz, or a factor of 4000 higher than the vibrational
frequency. For a 1200\,K grain near its sublimation temperature, the
1.66\,eV difference between the grain's binding energy and ionization
potential corresponds to an exponential suppression factor of
exp(-1.66\,eV/$k\times1200$\,K) = $10^{-7}$. Thus, for
$T\approx1200$\,K the frequency of thermionic emission will be only
$\sim4\times10^{-4}$ that of thermal sublimation of Fe, and can be
neglected.

Under interstellar conditions, there is a significant probability that
a Fe nanoparticle can be negatively charged: Fe$_n^-$.
The energy to remove an electron from an Fe$_n^-$ anion is equal
to the ``electron affinity'' EA of the neutral Fe$_n$.
Electron affinities have been determined from
photoelectron spectroscopy of iron clusters
\citep[][see Figure~\ref{fig:ea}]{Wang+Li+Zhang_2000}, with $EA \approx 2.5\,{\rm eV}$ for $n \approx 34$.
This is smaller than the estimated Fe atom binding energy
$B\approx 3.4\,{\rm eV}$ for $n\approx 34$.
It therefore seems likely that a hot Fe$_n^-$ will cool by
emitting an electron before an Fe atom is sublimed.

Under conditions typical of the cold neutral medium (CNM, see
Figure~\ref{fig:charge_dist}), we expect the majority of the Fe$_n$ to
be either neutral or positively charged. Accordingly, it does not
appear that cooling by thermionic emission will significantly reduce
the rate of sublimation of Fe atoms from Fe$_n$ clusters in the CNM. However, in
highly-ionized environments where collisions with electrons are
frequent, the grains can acquire negative charges (see Figure~\ref{fig:charge_dist}). In these
environments, thermionic emission may allow the grains to persist to
somewhat smaller radii.

\subsection{Desorption of H}

H atoms arrive at the surface of an uncharged Fe$_n$ nanoparticle
at a rate
\begin{align}
\dot{N}({\rm H}\,{\rm arrival}) 
&= n({\rm H})\left(\frac{8kT}{\pi m_{\rm H}}\right)^{1/2}\pi a^2 \nonumber
\\
&= 2.9\times10^{-8} 
\left(\frac{n({\rm H})}{30\,{\rm cm}^{-3}}\right)
\left(\frac{T}{10^2\,{\rm K}}\right)^{1/2}
\left(\frac{n}{34}\right)^{2/3}
~~~,
\end{align}
with somewhat higher collision rates (because of the induced-dipole
interaction) for clusters that are charged.  This is much higher than
the rate of Fe desorptions:

\begin{equation}
\dot{N}({\rm Fe\,sublimation}) = 
1.1\times10^{-11}
\left(\frac{n}{34}\right)
\left(\frac{10^5\,{\rm yr}}{\tau_{\rm subl}}\right)
\,{\rm s}^{-1}
~~~.
\end{equation}
A fraction $f_{\rm chem}$ of arriving H atoms will stick to the grain
and be chemisorbed.
Chemisorption of H atoms on bcc Fe has been studied
experimentally and theoretically.
The tightest binding appears to be for the Fe(110) surface,
with a binding energy $2.86\pm0.05\,{\rm eV}$/atom
\citep{Jiang+Carter_2003}.  Note that H$_2$ formation is energetically
unfavorable relative to two H atoms bound to Fe.
Only after monolayer coverage is achieved would H$_2$ formation take place
on the surface.

The estimated binding energy $B_{\rm H}\approx 2.86\,{\rm eV}$ of a chemisorbed
H atom on Fe\,(100) is less than the estimated binding energy
$B\approx 3.4\,{\rm eV}$ for an Fe atom.  Thus a hot Fe$_n$ cluster with
chemisorbed H will desorb H atoms more rapidly than Fe.
Loss of one or more H atoms will cool the cluster enough that no Fe
atoms will be desorbed during that thermal spike.

For Fe$_n$ clusters with submonolayer coverage of H, desorption of
adsorbed H atoms could potentially cool the cluster and reduce
the loss of Fe atoms. The effects of adsorbed H on the temperature
probability distribution for Fe$_n$ clusters, and on their survival in
the ISM, will be the subject of future investigation. However, at this
time it seems evident that desorption of H will lead to a significant
reduction in the minimum size of the Fe cluster that can survive in
the diffuse ISM, and we will therefore consider cluster sizes as small
as $n=25\ (a=4.0$\,\AA).

\subsection{Other Destruction Processes}
While we have emphasized destruction due to heating from ambient
starlight, other mechanisms may also limit the lifetimes of iron
nanoparticles in the ISM. In this section, we discuss the effects of
X-ray absorption and reactive sputtering.

Upon absorption of a $\sim1\,$keV X-ray, an iron nanoparticle would
likely undergo photoelectric emission of an L shell electron, followed
by emission of several Auger electrons. Some of the
remaining energy could conceivably evaporate one or more Fe atoms from
the grain surface. However, for an X-ray density of
$10^{-8}$\,cm$^{-3}$ \citep[see][Figure~12.1]{Draine_2011} and an
absorption efficiency $Q_{\rm abs} \simeq 1.4\times10^{-2}$ for a
1\,nm metallic iron grain, an X-ray absorption would occur only every
$3\times10^{5}$ years.

Chemisputtering, particularly due to H and O atoms, is another another
potential destruction pathway.  The surface chemistry, including the
effects of vacuum ultraviolet radiation, is poorly understood -- it is
conceivable, for instance, that impinging O atoms may be chemisorbed,
forming an oxide coating on the Fe nanoparticle, but also plausible
that such chemisorbed O atoms could be removed via chemisputtering by
impinging H atoms. Whether interstellar
chemistry is capable of maintaining a population of pure metallic Fe
nanoparticles is beyond the scope of this work, particularly in light
of the present uncertainties. Rather, we focus instead on the
observable consequences should such a population exist, and compute
the lifetime of such a grain exposed to the interstellar radiation
field.

\section{Grain Charge Distribution}
\label{sec:charge}
The grain charge distribution is the result of statistical equilibrium between the
processes of collisional charging and
photoelectric emission. Denoting the electron and ion collision rates
as $J_{\rm e}(Z,a)$ and $J_{\rm i}(Z,a)$, respectively, and the rate of
photoelectric emission $J_{\rm pe}(Z,a)$, in statistical
equilibrium we have

\begin{equation}
\label{eq:grain_charge}
\left[J_{\rm i}(Z,a) + J_{\rm pe}(Z,a)\right]f_a(Z) = J_{\rm
  e}(Z+1,a)f_a(Z+1)
~~~,
\end{equation}
where $f_a(Z)$ denotes the probability of a grain with radius $a$
having charge $Z$. To compute these rates for iron nanoparticles, we follow closely the
treatment of \citet{Weingartner+Draine_2001b} for carbonaceous and
silicate grains, adapting it where necessary to the case of metallic
iron. We refer the interested reader to that work for a more thorough
explication of the physics and derivations of the results employed
here. The charge distribution is calculated for three idealized ISM
phases-- the cold neutral medium (CNM), warm neutral medium (WNM), and
warm ionized medium (WIM). We adopt the parameterizations of these
media suggested by \citet{Draine+Lazarian_1998b} and presented in
Table~\ref{table:ism}.

\begin{deluxetable}{lccc}
      \tablecaption{Idealized ISM Phases \label{table:ism}}
    \tablehead{& \colhead{CNM} & \colhead{WNM} & \colhead{WIM}}
    \startdata
    $n_{\rm H}\,({\rm cm}^{-3})$ & 30 & 0.4 & 0.1 \\
    $T_g\,({\rm K})$ & 100 & 6000 & 8000 \\
    $x_{\rm H} \equiv n({\rm H}^+)/n_{\rm H}$ & 0.0012 & 0.1 & 0.99 \\
    $x_{\rm C} \equiv n({\rm C}^+)/n_{\rm H}$ & 0.0003 & 0.0003 & 0.001
    \enddata
    \tablecomments{Hydrogen number density $n_{\rm H}$, gas
      temperature $T_g$, H$^+$
      abundance $x_{\rm H}$, and C$^+$ abundance $x_{\rm C}$ adopted for the
      cold neutral medium (CNM), warm neutral medium (WNM), and warm ionized
      medium (WIM) following \citet{Draine+Lazarian_1998b}.}
\end{deluxetable}

\subsection{Collisional Charging}
The rate of sticking collisions of a species $i$ (in this work, we will
consider electrons as well as H$^+$ and C$^+$ ions) with a spherical
grain of radius $a$ and charge $Z$ is given by

\begin{equation}
J(Z,a) = n_i s_i(Z)\left(\frac{8kT_g}{\pi
    m_i}\right)^{1/2}\pi a^2\widetilde{J}(\tau_i,\nu_i)
~~~,
\end{equation}
where $n_i$ is the number density of species $i$, $m_i$ is the mass of
species $i$, and $T_g$ is the gas temperature. The function $s_i(Z)$
is the sticking efficiency of species $i$ onto a grain of charge $Z$,
and the function $\widetilde{J}(\tau,\nu)$ accounts for the
effect of Coulomb
focusing. It is given by \citep{Draine+Sutin_1987}

\begin{numcases}{\widetilde{J}(\tau,\nu) \simeq}
     1 + \left(\frac{\pi}{2\tau}\right)^{1/2},  & $\nu = 0$ \nonumber \\
     \left[1 - \frac{\nu}{\tau}\right]\left[1 + \left(\frac{2}{\tau -
           2\nu}\right)^{1/2}\right],  & $\nu < 0$  \nonumber \\
     \left[1 + \left(4\tau + 3\nu\right)^{-1/2}\right]^2{\rm
       exp}\left(-\theta_\nu/\tau\right), & $\nu > 0$, \nonumber \\
\end{numcases}
where $\tau_i = akT_g/q_i^2$ is the ``reduced temperature,'' $\nu_i =
Ze/q_i$, $q_i$ is the charge of species $i$, and

\begin{equation}
\label{eq:theta_nu}
\theta_\nu(\nu > 0) \simeq \frac{\nu}{1 + \nu^{-1/2}}
~~~.
\end{equation}

The grain composition enters into the charging
rates only through the sticking efficiency $s_i(Z)$. Unfortunately,
there is a dearth of laboratory data constraining the sticking
efficiencies of iron nanoparticles, so we adopt the empirical
treatment of \citet{Weingartner+Draine_2001b}. For ions, we assume
$s_{\rm ion} = 1$. For electrons, we adopt

\begin{equation}
\label{eq:stick}
s_e(Z > Z_{\rm min}) = 0.5\left(1 - e^{-a/l_e}\right)
~~~,
\end{equation}
where $l_e$ is the electron escape length (see
Section~\ref{subsec:photo_e}) and $Z_{\rm min}$ is the minimum grain
charge (see Equation~\ref{eq:zmin}). For $Z = Z_{\rm min}$, the grain
will autoionize following an electron collision, and so $s_e(Z_{\rm
  min}) = 0$. Note that in the absence of
experimental data, we have ignored suppression of the
sticking efficiency in the smallest grains
\citep[cf.][Equation~28]{Weingartner+Draine_2001b}.

\begin{figure*}
    \centering
        \scalebox{1.0}{\includegraphics{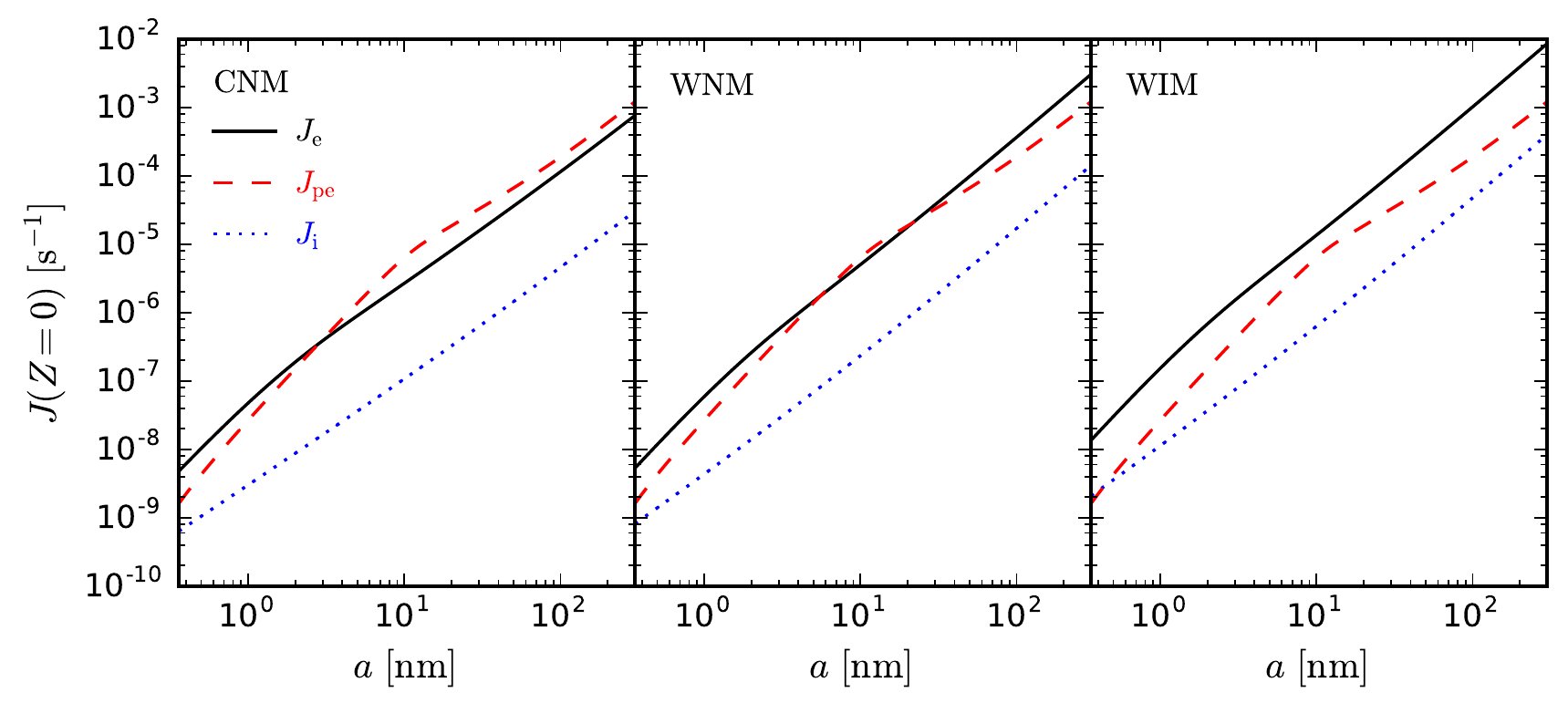}}  
    \caption{The rate of collisional charging due to electrons (black
      solid) and ions (blue dotted) and the rate photoelectric
      charging (red dashed) are plotted as a function of grain size
      for CNM (left), WNM (middle) and WIM (right) environments for
      neutral iron grains. The relatively inefficient photoelectric
      yield of these grains can cause electrons to accumulate on the
      grains leading to preferentially negative grain charges.} \label{fig:charging_rate} 
\end{figure*}

In Figure~\ref{fig:charging_rate}, we plot the collisional charging rates for iron
nanoparticles as a function of size in each of the idealized ISM
environments detailed in Table~\ref{table:ism}.

\subsection{Photoelectric Emission}
\label{subsec:photo_e}
We will now compute the rate of photoelectric emission of iron
nanoparticles as a function of grain size and charge. Consider spherical iron nanoparticles with mass density $\rho
= 7.87$\,g\,cm$^{-3}$. The number of iron atoms $n$ in a cluster with
radius $a$ can be approximated by

\begin{equation}
\label{eq:n_atoms}
n = 352 \left(\frac{a}{\rm 10\,\AA}\right)^3
~~~.
\end{equation}

\begin{figure}
    \centering
        \scalebox{1.0}{\includegraphics{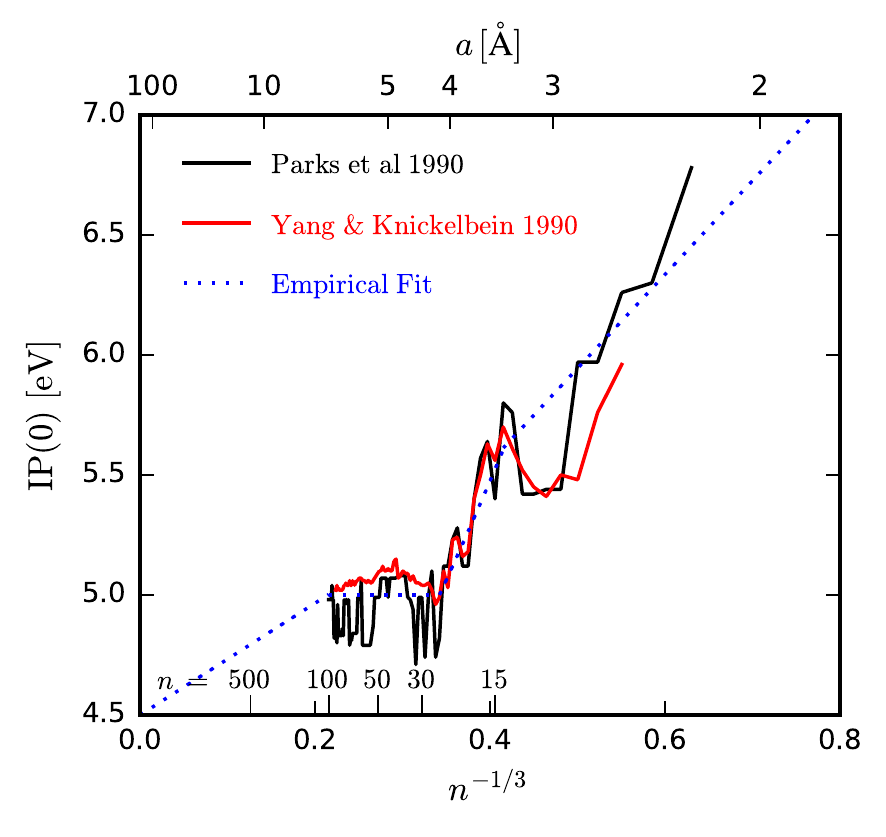}}  
    \caption{The ionization potential (IP) for clusters of $n$ iron
      atoms as determined by \citet{Parks+Klots+Riley_1990} (black)
      and \citet{Yang+Knickelbein_1990} (red). Agreement between the
      two sets of experimental data is good; however, the data are in
      sharp disagreement with the spherical drop model. We overlay our
    empirical fit to the experimental data (blue dotted).} \label{fig:ip} 
\end{figure}

The ionization potential (IP, also referred to as the ionization
energy) can be calculated classically with the ``spherical drop
model'' (SDM) in which the cluster is modeled as a conducting
sphere. In the SDM, the
energy required to move an electron from the surface of a sphere of
charge $Z$ is given by

\begin{equation}
\label{eq:spherical_drop}
{\rm IP}_{\rm SDM}(Z) = W + \left(Z + \frac{1}{2}\right)\frac{e^2}{a}
~~~,
\end{equation} 
where $W$ is the bulk work function of the material. For metallic
iron, $W = 4.5\,{\rm eV}$ \citep{Eastman_1970}. In contrast to the
predictions of the spherical drop model, experimental data
indicate very little evolution of IP between $30 \lesssim n \lesssim
100$ \citep{Parks+Klots+Riley_1990, Yang+Knickelbein_1990}. We plot
these data in Figure~\ref{fig:ip}.

We approximate IP(0) with an empirical fit to the experimental data
(see Figure~\ref{fig:ip}) based on a modification of
Equation~\ref{eq:spherical_drop}. For $Z \geq 0$, we take

\begin{numcases}{{\rm IP}(Z \geq 0) =}
      3.99\,{\rm eV} + (Z + 0.38)\frac{e^2}{a},  & $n < 14$ \nonumber
      \\
      2.22\,{\rm eV} + (Z+0.80)\frac{e^2}{a},  & $14 \leq n < 25$
      \nonumber \\
      5.00\,{\rm eV} + Z\frac{e^2}{a},  & $25 \leq n < 100$ \nonumber \\
      W + (Z + 0.23)\frac{e^2}{a},  & $n \geq 100$ \label{eq:ip}
\end{numcases}

Grains with $Z < 0$ have ``extra'' electrons in their molecular
orbitals. The energy required to ionize a grain of charge $Z < 0$ is
equal to the energy gained by adding an electron to a grain of charge
$Z + 1$. This latter quantity is called the ``electron affinity''
EA, and so IP($Z<0$) = EA($Z+1$). In the case of iron, which should not
have an energy band gap, the EA is also expected to follow the SDM.

\begin{figure}
    \centering
        \scalebox{1.0}{\includegraphics{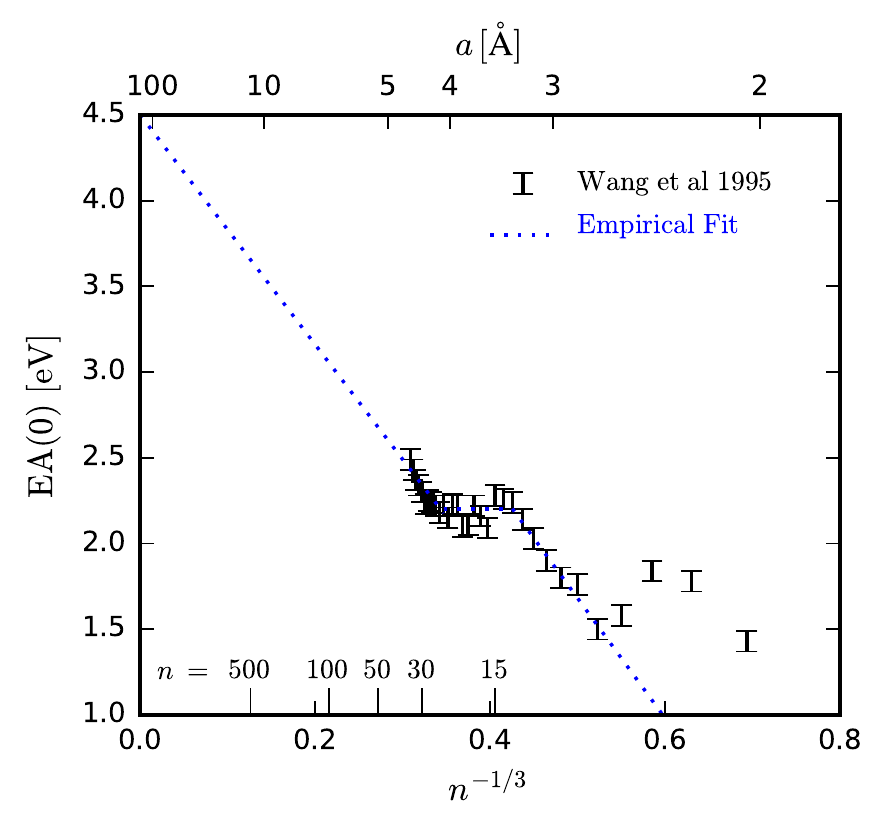}}  
    \caption{The electron affinity for neutral iron clusters of $n$
      atoms as measured by \citet{Wang+Cheng+Fan_1995}. We overlay our
      model (Equation~\ref{eq:ea}) which extrapolates the experimental
      data to the bulk limit with a modified form of the spherical
      drop model (blue dotted line).} \label{fig:ea} 
\end{figure}

\citet{Wang+Cheng+Fan_1995} employed photoelectron spectroscopy to
measure the EA for iron clusters with $3
\leq n \leq 34$ atoms and found that EA scales with $n^{-1/3}$ as expected
from the SDM above $n \simeq 20$. We find good numerical agreement with
their results, as shown in Figure~\ref{fig:ea}, by adopting

\begin{numcases}{{\rm EA}(Z) =}
      5.18\,{\rm eV} + (Z - 0.69)\frac{e^2}{a},  & $n <
      13$ \nonumber
      \\
      2.20\,{\rm eV} + Z\frac{e^2}{a},  & $13 \leq n < 24$ \nonumber
      \\
      W + (Z - 0.66)\frac{e^2}{a},  & $n \geq 24$ \label{eq:ea}
\end{numcases}
which, with Equation~\ref{eq:ip},
fully specifies IP as a function of $a$ and $Z$.

Let us define $h\nu_{\rm pet}$ as the threshold photon energy for
photoionization. For grains with $Z \geq -1$, this
energy is simply equal to the energy difference IP required to move an
electron from the grain to infinity.  However, when $Z < -1$ the grain
repels the escaping electron, so that it necessarily arrives at
infinity with kinetic energy $\geq E_{\rm min}$.  Thus the minimum photon
energy to eject the electron must be

\begin{numcases}{h\nu_{\rm pet}(Z,a) =}
     {\rm IP}(Z,a),  & $Z \geq -1$ \nonumber
      \\
     {\rm IP}(Z,a) + E_{\rm min}(Z,a),  & $Z < -1$ 
~~~.
\end{numcases}
We adopt the $E_{\rm min}$ prescription of \citet{vanHoof+etal_2004}:

\begin{equation}
E_{\rm min} =
\theta_\nu\left(|Z+1|\right)\left[1-0.3\left(\frac{a}{10\,{\rm
        \AA}}\right)^{-0.45}|Z+1|^{-0.26}\right]
~~~,
\end{equation}
where $\theta_\nu$ is given by Equation~\ref{eq:theta_nu}.

We now quantify the expected number of electrons yielded by a photon
of energy $h\nu > h\nu_{\rm pet}$. Consider first the yield $y_0'$ of
electrons that escape the grain surface but may or may not escape to
infinity. We employ the model

\begin{equation}
y_0'\left(h\nu, Z, a\right) = y_0\left(\Theta\right)
y_1\left(h\nu,a\right)
~~~,
\end{equation}
where $y_0(\Theta)$ primarily governs the frequency-dependence of the
yield in the limit of bulk material, $y_1$ quantifies the yield enhancement due to geometrical
effects, and $\Theta$ is a parameter defined by

\begin{numcases}{\Theta =}
    h\nu - h\nu_{\rm pet} + \left(Z+1\right)e^2/a,  & $Z \geq 0$ \nonumber
      \\
    h\nu - h\nu_{\rm pet},  & $Z < 0$ \label{eq:Theta}
~~~.
\end{numcases}

$y_1$ can be approximated by \citep{Draine_1978}

\begin{equation}
y_1 = \left(\frac{\beta}{\alpha}\right)^2\frac{\alpha^2 - 2\alpha + 2
  - 2e^{-\alpha}}{\beta^2 - 2\beta + 2 - 2e^{-\beta}}
~~~,
\end{equation}
where $\alpha = a/l_a + a/l_e$, $\beta = a/l_a$, $l_e$ is the electron
escape length, and $l_a$ is the photon attenuation length. For a
material with complex refractive index $m\left(\lambda\right)$, $l_a$
is given by

\begin{equation}
l_a = \frac{\lambda}{4\pi\, {\rm Im}\left(m\right)}
~~~.
\end{equation}
We employ the complex refractive index for metallic Fe from
\citet{Draine+Hensley_2013}.

The electron escape length $l_e$ is the mean free path of an electron
to inelastic scattering and is a parameter in models for the
photoelelectric yield of grains as well as for the sticking efficiency
(see Equation~\ref{eq:stick}). The energy-dependence of $l_e$ has been measured for a
range of materials and is generally found to decrease with electron
energy at low energy, have a minimum between 10 and 100\,eV above the
Fermi surface, then increase to high energies
\citep{Seah+Dench_1979}. As experimental data are sparse and as the
minimum is coincident with energies of astrophysical interest, some
previous studies have adopted an energy-independent $l_e = 10\,$\AA\
\citep{Bakes+Tielens_1994,
  Weingartner+Draine_2001b}. \citet{Kimura_2016} instead adopts the
empirical prescription of \citet{Seah+Dench_1979} for the case of pure
elements:

\begin{equation}
l_e = 143\left(\frac{E}{\rm eV}\right)^{-2} + 0.054\left(\frac{E}{\rm
    eV}\right)^{1/2}\ {\rm nm}
~~~,
\end{equation}
where here $E$ is the energy above the Fermi level. This prescription
yields $l_e = 16\,$\AA\ at $E = 10$\,eV. Given the current
state of experimental uncertainty, as well as the rough agreement
between approaches in the regime of interest, we opt for the simpler
energy-independent treatment with $l_e = 10\,$\AA.

\begin{figure}
    \centering
        \scalebox{1.0}{\includegraphics{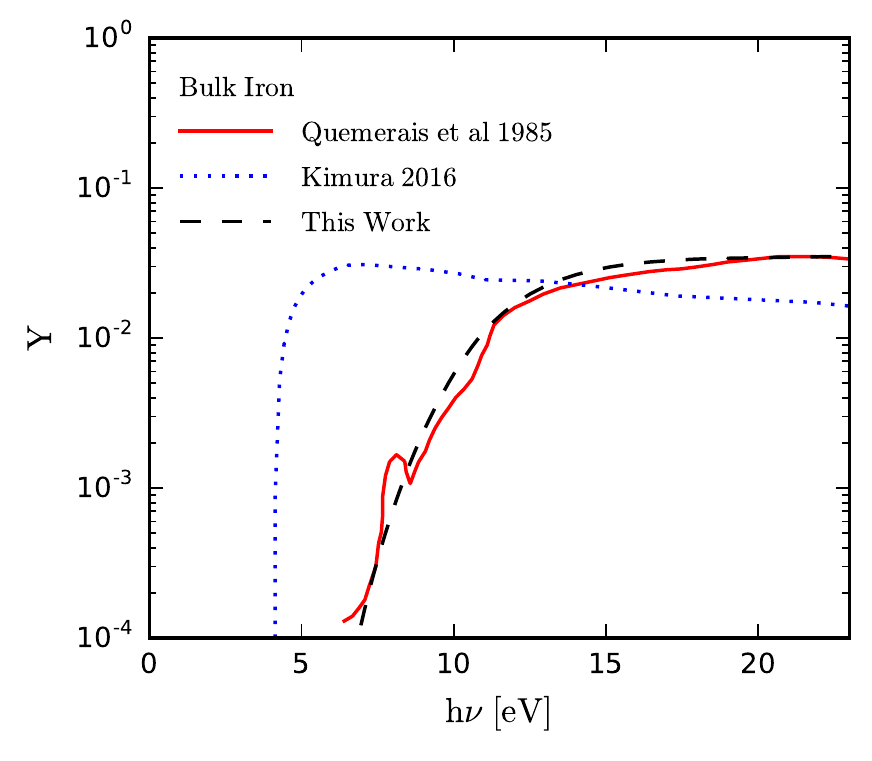}}  
    \caption{The photoelectric yield $Y$ for bulk iron as measured by
      \citet{Quemerais+etal_1985} (red solid line) and for our model with
      $a = 100\,$nm (black dashed line). We also plot the model of
      \citet{Kimura_2016} (blue dotted), but note that, unlike our
      model, it was not explicitly calibrated on the experimental
      data.} \label{fig:y_fe} 
\end{figure}

Previous studies \citep{Bakes+Tielens_1994,
  Weingartner+Draine_2001b} have determined $y_0$ by matching model
predictions to experimental data at some fixed grain size and
charge. In Figure~\ref{fig:y_fe}, we plot the bulk photoelectric yield
of metallic iron as measured by \citet{Quemerais+etal_1985}. We find
that we can obtain reasonable agreement with the experimental
data by adopting:

\begin{equation}
y_0 = \frac{2.1\times10^{-3}\left(\Theta/W\right)^5}{1 +
  6.8\times10^{-3}\left(\Theta/W\right)^5}
~~~,
\end{equation}
with $\Theta$ given by Equation~\ref{eq:Theta}.

Finally, we define $y_2$ as the fraction of electrons that emerge from
the grain surface that escape to infinity. We follow
\citet{Weingartner+Draine_2001b} in assuming a parabolic distribution
of electron energies, yielding

\begin{figure}
    \centering
        \scalebox{1.0}{\includegraphics{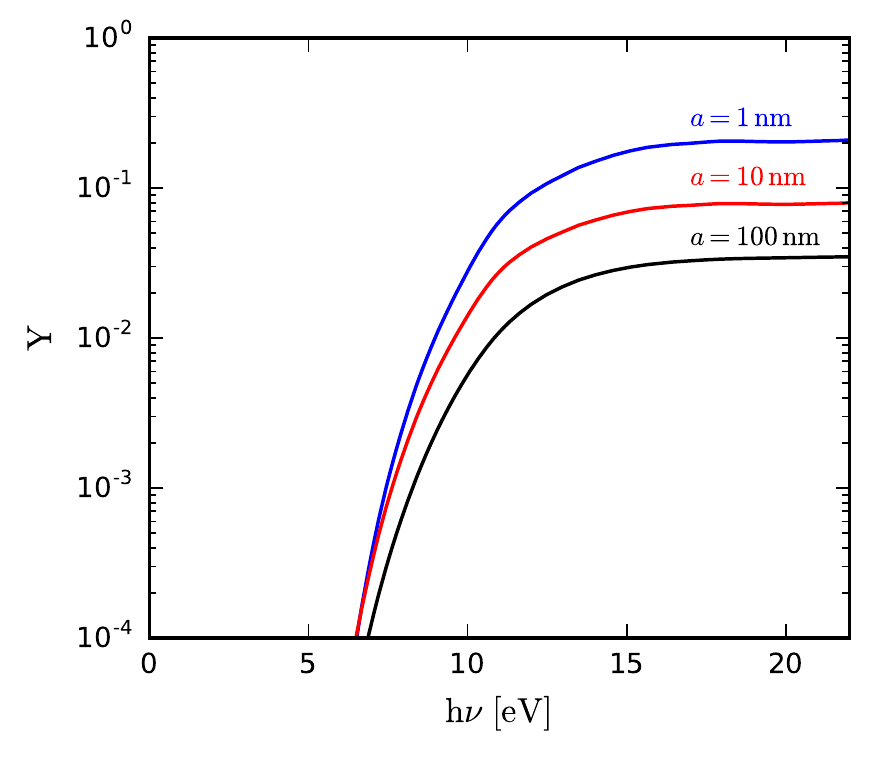}}  
    \caption{The photoelectric yield of neutral iron nanoparticles of radius
      1, 10, and 100\,nm. The enhancement of the
      photoelectric yield due to small particle effects saturates at
      roughly $a = 1\,$nm, so the energy-dependent photoelectric yield
      of the $a = 1$\,nm grains is also representative of smaller grains.} \label{fig:y_size} 
\end{figure}

\begin{numcases}{y_2 =}
   E_{\rm high}^2\left(E_{\rm high} - 3E_{\rm low}\right)/\left(E_{\rm
       high}-E_{\rm low}\right)^3,  & $Z \geq 0$ \nonumber
      \\
    1,  & $Z < 0$, \nonumber \\
\end{numcases}
where $E_{\rm low} = -(Z+1)e^2/a$ and $E_{\rm high} = h\nu  -
h\nu_{\rm pet}$. Thus, the photoelectric yield $Y$ is given by

\begin{equation}
Y(h\nu,Z,a) = y_2\, {\rm min}\left(y_0', 1\right)
~~~.
\end{equation}
In Figure~\ref{fig:y_size} we plot the yield for iron nanoparticles of
radius 1, 10, and 100\,nm. The yield for particles with $a <
1$\,nm is similar to that of 1\,nm grains.

As a final step in computing the photoelectric charging rate $J_{\rm pe}$, we consider the
photodetachment of electrons above the
valence band for grains with $Z < 0$. The photodetachment threshold
energy $h\nu_{\rm pdt}$ is

\begin{equation}
h\nu_{\rm pdt} = {\rm EA}\left(Z+1,a\right) + E_{\rm
  min}\left(Z,a\right)
~~~.
\end{equation}
We follow \citet{Weingartner+Draine_2001b} in adopting

\begin{equation}
\sigma_{\rm pdt} = \frac{-1.2\times10^{-17}Zx}{\left(1 + x^2/3\right)^2}\ {\rm
      cm}^2
\end{equation}
for the photodetachment cross section $\sigma_{\rm pdt}$, where $x
\equiv \left(h\nu - h\nu_{\rm pdt}\right)/3\,{\rm eV}$.

For a spherical grain of radius $a$ with absorption efficiency $Q_{\rm
  abs}\left(\nu\right)$ in a radiation field with specific energy
density $u_\nu\left(\nu\right)$, the photoemission rate is

\begin{equation}
J_{\rm pe}\left(Z,a\right) = \pi a^2 \int_{\nu_{\rm pet}}^{\infty}
{\rm d}\nu\, Y\, Q_{\rm abs}\,\frac{cu_\nu}{h\nu} + \int_{\nu_{\rm
    pdt}}^{\infty} {\rm d}\nu\, \sigma_{\rm pdt}\,\frac{cu_\nu}{h\nu}
~~~.
\end{equation}
We adopt the absorption efficiencies of metallic Fe computed by
\citet{Draine+Hensley_2013} and the starlight energy density spectrum
of \citet{Mathis+Mezger+Panagia_1983}. The rate of photoelectric
emission is compared to the collisional charging rates in
Figure~\ref{fig:charging_rate} for iron grains of various sizes in
CNM, WNM, and WIM environments.

\subsection{Charge Distribution}
Once the collisional charging rates and photoelectric emission rates
have been computed, Equation~\ref{eq:grain_charge} can be solved
recursively after a minimum and maximum grain charge ($Z_{\rm min}$ and
$Z_{\rm max}$, respectively) have been determined. 

If an electron is added to a grain of sufficiently negative potential,
the grain will autoionize. Denoting the threshold autionization
threshold potential $U_{\rm ait}$, we have

\begin{equation}
\label{eq:zmin}
Z_{\rm min} = 1 + \left \lfloor{\frac{U_{\rm ait}}{e/a}}\right \rfloor
~~~,
\end{equation}
where $\left\lfloor{}\right\rfloor$ is the floor function. As graphite
and iron have similar work functions (4.4 and 4.5\,eV, respectively),
we adopt the $U_{\rm ait}$ derived for carbonaceous grains by
\citet{Weingartner+Draine_2001b}: 

\begin{equation}
\frac{-U_{\rm ait}}{\rm V} = 3.9 + 0.12\left(a/{\rm \AA}\right) +
2\left({\rm \AA}/a\right)
~~~.
\end{equation}

A grain achieves its maximum charge when its ionization potential
becomes larger than the maximum photon energy of the ambient radiation field
$h\nu_{\rm max}$, here
taken to be 13.6\,eV. Thus,

\begin{numcases}{Z_{\rm max} =}
     \left \lfloor{0.62 + \frac{a}{e^2}\left(h\nu_{\rm max} - 3.99\,{\rm
             eV}\right)}\right \rfloor,  & $n <
     14$ \nonumber
      \\
     \left \lfloor{0.20 + \frac{a}{e^2}\left(h\nu_{\rm max} - 2.22\,{\rm
             eV}\right)}\right \rfloor,  & $14 \leq n <
     25$ \nonumber \\
     \left \lfloor{1 + \frac{a}{e^2}\left(h\nu_{\rm max} - 5\,{\rm
             eV}\right)}\right \rfloor,  & $25 \leq n <
     100$ \nonumber \\
      \left \lfloor{0.77 + \frac{a}{e^2}\left(h\nu_{\rm max} -
            W\right)}\right \rfloor,  & $n \geq
      100$. \label{eq:zmax}
\end{numcases}

\begin{figure*}
    \centering
        \scalebox{1.0}{\includegraphics{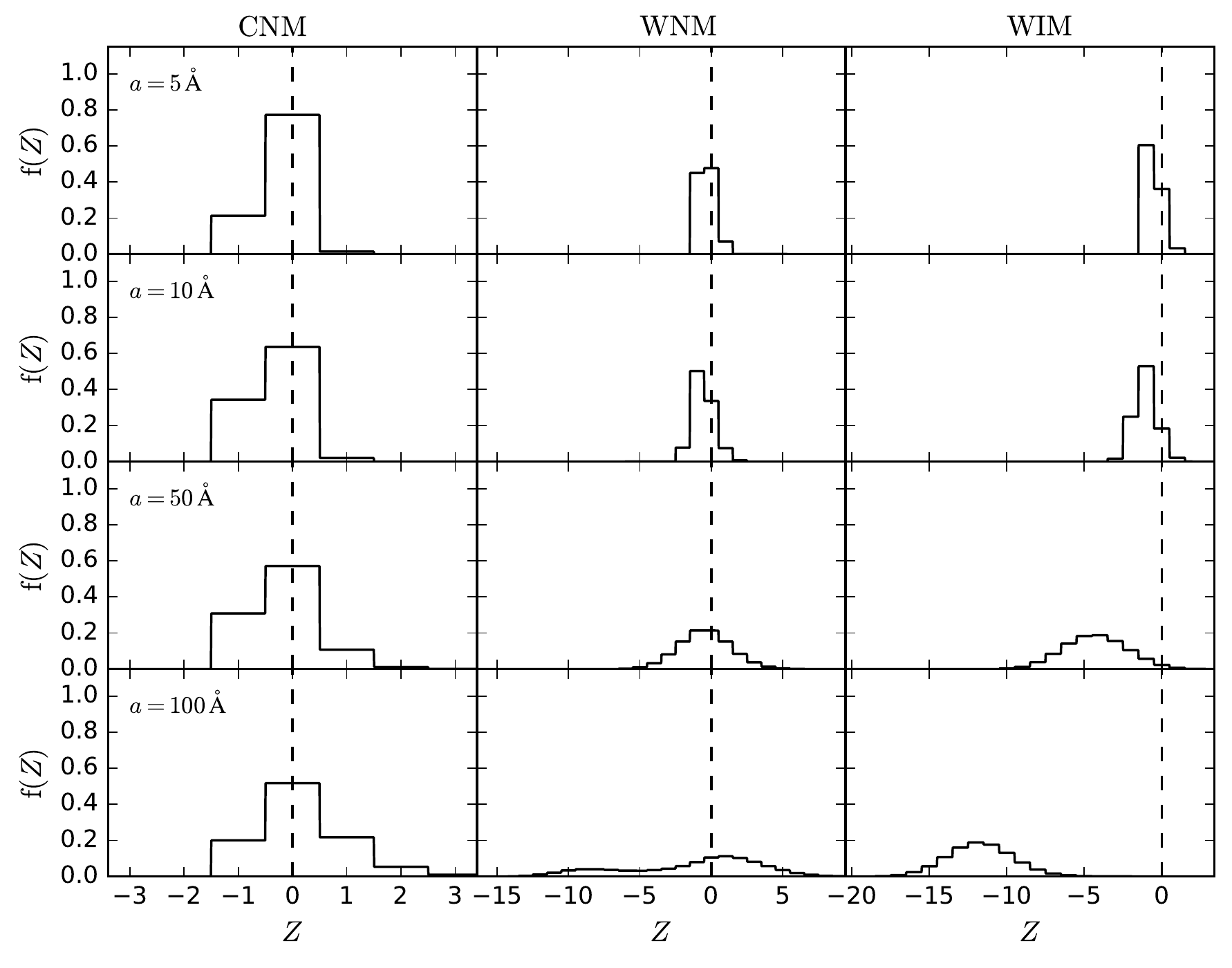}}  
    \caption{The grain charge distribution for 5 (top row), 10
      (second row), 50 (third row), and 100\,\AA\ (bottom row) iron
      nanoparticles
      in the CNM (left column), WNM (middle column), and WIM (right
      column). Of particular note is the tendency for these grains to
      be charged to {\it negative} potentials, particularly in the WNM
    and WIM due to their relatively low photoelectric emission rates
    (see Figure~\ref{fig:charging_rate}).} \label{fig:charge_dist} 
\end{figure*}

In Figure~\ref{fig:charge_dist} we present the final charge
distributions of iron nanoparticles in the CNM, WNM, and WIM for
grain sizes of 5, 10, 50, and 100\,\AA. 

Perhaps the most striking feature of the grain charge distributions is
the tendency of iron nanoparticles to acquire {\it negative} charge,
in contrast to the typically positively-charged carbonaceous and
silicate grains \citep[see][Figure~10]{Weingartner+Draine_2001b}. This is due largely to the
relatively inefficient photoelectric yield of the iron
grains.

\section{Photoelectric Heating}
\label{sec:photo_e}
Energetic electrons photoelectrically ejected from grains serve as a
major heating source of the ambient interstellar gas. In this section,
we quantify the potential contribution of iron nanoparticles to the
total photoelectric heating from dust. We again follow closely
\citet{Weingartner+Draine_2001b}.

As discussed above, electrons are liberated from grains via
photoemission of valence electrons and the photodetachment of
electrons above the valence band. Let $\Gamma'_{\rm pe}$ be the
total heating rate per grain from photoelectric emission, which can be
computed for iron nanoparticles from \citet[][Equations 38 -
40]{Weingartner+Draine_2001b} by employing the model of photoelectric
emission from iron grains presented in this work.

\begin{figure}
    \centering
        \scalebox{1.0}{\includegraphics{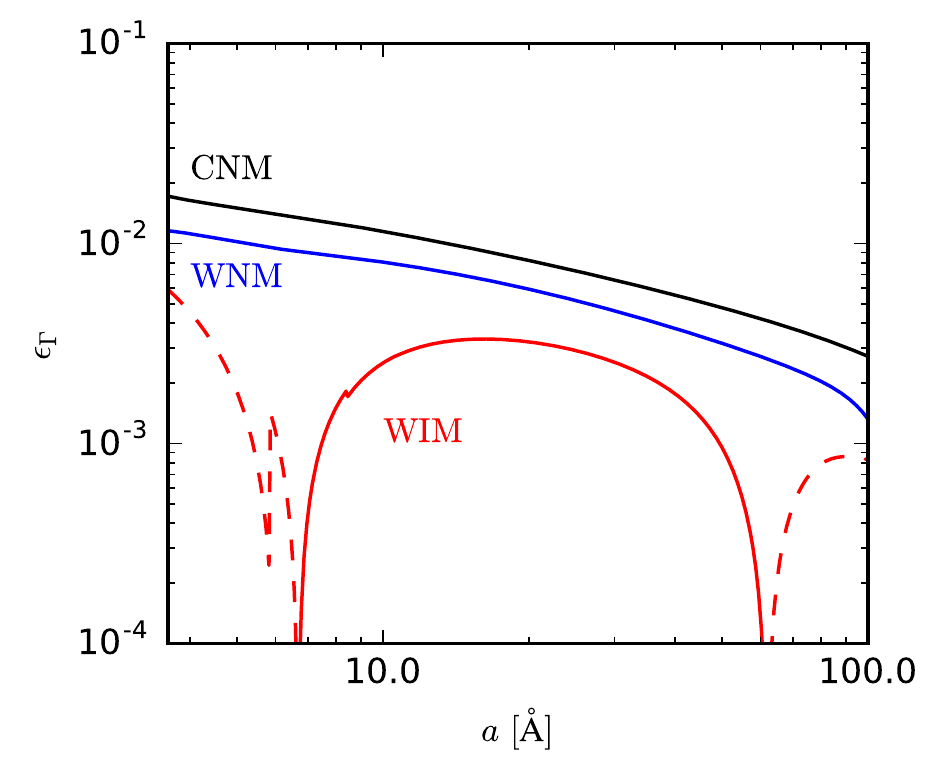}}  
    \caption{The photoelectric heating efficiency $\epsilon_\Gamma$ as
    a function of grain size in the CNM, WNM, and WIM. For very small
    and large grains in the WIM, collisional cooling is more efficient
  than photoelectric heating, giving these grains a net {\it cooling}
  effect on the gas ($\epsilon_\Gamma < 0$ is indicated with a dashed
  line). As was found for carbonaceous and silicate
  grains, photoelectric heating from iron grains is dominated by the
  smallest grains.} \label{fig:photoheating} 
\end{figure}

Grains also capture electrons and thereby remove
energy from the gas, and so the total heating rate per grain is
lessened by the energy removal rate $\Lambda_{\rm gr}'$
\citep[see][Equation 42]{Weingartner+Draine_2001b}. The total gas
heating efficiency per grain $\epsilon_\Gamma$ is defined by

\begin{equation}
\epsilon_\Gamma = \frac{\Gamma_{\rm pe}' - \Lambda_{\rm gr}'}{\pi a^2
  c \int u_\nu\,Q_{\rm abs}\ {\rm d}\nu}
~~~.
\end{equation}
We plot $\epsilon_\Gamma$ for iron nanoparticles in various
interstellar environments in Figure~\ref{fig:photoheating}. As
expected due to the enhancement of the photoelectric yield in small
particles, the smallest grains are most effective at heating the
gas. The smallest iron nanoparticles are only slightly less efficient at
heating the gas than silicate grains and approximately a factor of two
less efficient than carbonaceous grains as computed by
\citet{Weingartner+Draine_2001b}.

It is not known how much of the interstellar iron is in the form of
free-flying iron nanoparticles. Adopting a solid-phase interstellar
iron abundance of 41\,ppm \citep{Bensby+etal_2005, Lodders+etal_2009,
  Jenkins_2009} and a fraction $Y_{\rm Fe}$ of that iron in the form
of free-flying iron nanoparticles, we assume a log-normal size
distribution for $a > a_{\rm min}$:

\begin{equation}
\label{eq:size_dist}
\frac{1}{n_{\rm H}} \frac{{\rm d}n_{\rm Fe}}{{\rm d}a} =
  \frac{A}{a}{\rm exp}\left\{-\frac{1}{2}\left[\frac{{\rm
            ln}\left(a/a_0\right)}{\widetilde{\sigma}}\right]^2\right\}
~~~,
\end{equation}
where the parameters $a_0$ and $\widetilde{\sigma}$ determine the peak size and
width of the size distribution. $A$ is a normalization constant
given by

\begin{align}
A &= \frac{3}{\left(2\pi\right)^{3/2}}\frac{{\rm
    exp}\left(-4.5\widetilde{\sigma}^2\right)}{a_0^3\rho\widetilde{\sigma}} \times \nonumber
  \\ 
& \frac{m_{\rm Fe}b_{\rm Fe}}{1 + {\rm
  erf}\left[3\widetilde{\sigma}/\sqrt{2} + {\rm ln}\left(a_0/a_{\rm min}\right)/\widetilde{\sigma}\sqrt{2}\right]}
\end{align}
for $a > a_{\rm min}$ and zero otherwise, where $\rho = 7.87\,{\rm
  g}\,{\rm cm}^{-3}$ is the grain mass density, $m_{\rm Fe } = 9.27\times10^{-23}\,$g is
the mass of an iron atom, and $b_{\rm Fe} = 41 Y_{\rm Fe}\times
10^{-6}$ is the number of Fe atoms per H consumed by this grain
population. We adopt $a_0 = a_{\rm min} = 4.5\,$\AA\ and $\widetilde{\sigma} =
0.3$ as an illustrative case that roughly maximizes the photoelectric
heating of this population by concentrating the grain mass at the
smallest viable sizes.

With this size distribution, we compute the total photoelectric
heating rate per H nucleon per Habing flux via

\begin{equation}
\frac{\Gamma_{\rm tot}}{G n_{\rm H}} = \int \frac{\Gamma_{\rm pe}' -
  \Lambda_{\rm gr}'}{G} \frac{1}{n_{\rm H}} \frac{{\rm d}n_{\rm
    Fe}}{{\rm d}a}\ {\rm d}a
\end{equation}
where the adopted radiation field of
\citet{Mathis+Mezger+Panagia_1983} has G = 1.13.

This population of iron nanoparticles yields a total heating rate per
H nucleon per
Habing flux of $2\times10^{-26}Y_{\rm Fe}$ and
$1\times10^{-26}Y_{\rm Fe}$\,erg\,s$^{-1}$\,H$^{-1}$ in the CNM and WNM,
respectively. In
contrast, \citet{Weingartner+Draine_2001b} found heating rates of
$6\times10^{-26}$ and $3\times10^{-26}$\,erg\,s$^{-1}$\,H$^{-1}$ in
the CNM and WNM, respectively, for a combination of
carbonaceous and silicate grains. In the WIM, the iron grains provide
a small source of {\it cooling} at the level of
$2\times10^{-27}Y_{\rm Fe}$\,erg\,s$^{-1}$\,H$^{-1}$, but this is
small relative to the heating rates that have been estimated for
photoionization from starlight or photoelectric heating from dust
\citep[$\sim
10^{-25}-10^{-24}$\,erg\,s$^{-1}$\,H$^{-1}$][]{Mathis_2000,
  Weingartner+Draine_2001b}.

Iron nanoparticles therefore appear able to make a minor but
non-negligible contribution to the total photoelectric heating
provided some ($Y_{\rm Fe} \gtrsim 10\%$) of the interstellar iron
exists in the form of small, free-flying particles.

\section{Conclusions}
\label{sec:conclusion}
The abundance of solid phase iron in the interstellar medium suggests
that iron nanoparticles may be a substantial component of interstellar
dust. Such particles could contribute to the interstellar extinction
and to the infrared emission. Magnetic dipole emission and rotational
emission from these grains could potentially contribute importantly to
the Galactic dust SED at submillimeter
and microwave wavelengths. In this work, we have investigated the
physics of these grains, focusing on their thermodynamics,
photodestuction, and charging. Our principal conclusions are as follows:

\begin{enumerate}
\item  Bare Fe$_n$ nanoparticles are unstable to photodestuction in the
   interstellar radiation field below $n=34\ (a=4.5$\,\AA).  In the presence
   of H atoms, Fe nanoparticles will be partially hydrogenated, and
   cooling by thermal desorption of H atoms will allow the Fe cores to
   survive down to smaller sizes, perhaps $n=25\ (a=4.0$\,\AA) in the diffuse
   ISM.
\item Due to a photoelectric yield lower than that of either carbonaceous or
  silicate grains, iron grains tend to be charged to negative
  potentials particularly in environments such as the WNM or WIM where
  the electron collision rate is high.
\item If $\gtrsim 10\%$ of the interstellar iron is in the form of
  metallic iron nanoparticles, these grains could increase the
  photoelectric heating from dust by up to tens of percent relative to
  models with only carbonaceous and silicate dust.
\item A large population of interstellar metallic iron nanoparticles
  can be considered a viable component of interstellar dust without
  having an unphysically large impact on interstellar heating.
\end{enumerate}

The results presented in this work rest, where possible, upon
available laboratory data. However, for materials of size and
composition of astrophysical interest, the data are sparse. Laboratory
measurements of the photoelectric yield of small iron clusters, as
well as carbonaceous and silicate materials, would
be of particular use, as would additional measurement of the
ionization potential and electron affinity for larger iron clusters
($n > 100$).

\acknowledgments
{We thank referee J. Nuth and a second anonymous referee for helpful
  comments. BTD acknowledges support from NSF grant
  AST-1408723. The research was carried out in part at the Jet Propulsion
  Laboratory, California Institute of Technology, under a contract
  with the National Aeronautics and Space Administration.}

\bibliography{/Users/bhensley/Dropbox/mybib}

\begin{thebibliography}{}
\expandafter\ifx\csname natexlab\endcsname\relax\def\natexlab#1{#1}\fi

\bibitem[{{Altobelli} {et~al.}(2016){Altobelli}, {Postberg}, {Fiege},
  {Trieloff}, {Kimura}, {Sterken}, {Hsu}, {Hillier}, {Khawaja},
  {Moragas-Klostermeyer}, {Blum}, {Burton}, {Srama}, {Kempf}, \&
  {Gruen}}]{Altobelli+etal_2016}
{Altobelli}, N., {Postberg}, F., {Fiege}, K., {et~al.} 2016, Science, 352, 312

\bibitem[{{Bakes} \& {Tielens}(1994)}]{Bakes+Tielens_1994}
{Bakes}, E.~L.~O., \& {Tielens}, A.~G.~G.~M. 1994, \apj, 427, 822

\bibitem[{{Bensby} {et~al.}(2005){Bensby}, {Feltzing}, {Lundstr{\"o}m}, \&
  {Ilyin}}]{Bensby+etal_2005}
{Bensby}, T., {Feltzing}, S., {Lundstr{\"o}m}, I., \& {Ilyin}, I. 2005, \aap,
  433, 185

\bibitem[{{Bradley}(1994)}]{Bradley_1994}
{Bradley}, J.~P. 1994, Science, 265, 925

\bibitem[{{Chlewicki} \& {Laureijs}(1988)}]{Chlewicki+Laureijs_1988}
{Chlewicki}, G., \& {Laureijs}, R.~J. 1988, \aap, 207, L11

\bibitem[{{Cox}(1990)}]{Cox_1990}
{Cox}, P. 1990, \aap, 236, L29

\bibitem[{{Davis} \& {Greenstein}(1951)}]{Davis+Greenstein_1951}
{Davis}, Jr., L., \& {Greenstein}, J.~L. 1951, \apj, 114, 206

\bibitem[{{Desai}(1986)}]{Desai_1986}
{Desai}, P.~D. 1986, Journal of Physical and Chemical Reference Data, 15, 967

\bibitem[{{Draine}(1978)}]{Draine_1978}
{Draine}, B.~T. 1978, \apjs, 36, 595

\bibitem[{{Draine}(2011)}]{Draine_2011}
---. 2011, {Physics of the Interstellar and Intergalactic Medium} (Princeton,
  NJ: Princeton University Press)

\bibitem[{{Draine} \& {Hensley}(2012)}]{Draine+Hensley_2012}
{Draine}, B.~T., \& {Hensley}, B. 2012, \apj, 757, 103

\bibitem[{{Draine} \& {Hensley}(2013)}]{Draine+Hensley_2013}
---. 2013, \apj, 765, 159

\bibitem[{{Draine} \& {Lazarian}(1998)}]{Draine+Lazarian_1998b}
{Draine}, B.~T., \& {Lazarian}, A. 1998, \apj, 508, 157

\bibitem[{{Draine} \& {Lazarian}(1999)}]{Draine+Lazarian_1999}
---. 1999, \apj, 512, 740

\bibitem[{{Draine} \& {Li}(2001)}]{Draine+Li_2001}
{Draine}, B.~T., \& {Li}, A. 2001, \apj, 551, 807

\bibitem[{{Draine} \& {Sutin}(1987)}]{Draine+Sutin_1987}
{Draine}, B.~T., \& {Sutin}, B. 1987, \apj, 320, 803

\bibitem[{{Duan} {et~al.}(2007){Duan}, {Ding}, {Ros{\'e}n}, {Harutyunyan},
  {Curtarolo}, \& {Bolton}}]{Duan+etal_2007}
{Duan}, H., {Ding}, F., {Ros{\'e}n}, A., {et~al.} 2007, Chemical Physics, 333,
  57

\bibitem[{{Duley}(1978)}]{Duley_1978}
{Duley}, W.~W. 1978, \apjl, 219, L129

\bibitem[{{Eastman}(1970)}]{Eastman_1970}
{Eastman}, D.~E. 1970, \prb, 2, 1

\bibitem[{{Guhathakurta} \& {Draine}(1989)}]{Guhathakurta+Draine_1989}
{Guhathakurta}, P., \& {Draine}, B.~T. 1989, \apj, 345, 230

\bibitem[{{Henning} {et~al.}(1995){Henning}, {Begemann}, {Mutschke}, \&
  {Dorschner}}]{Henning+etal_1995}
{Henning}, T., {Begemann}, B., {Mutschke}, H., \& {Dorschner}, J. 1995, \aaps,
  112, 143

\bibitem[{{Hensley} \& {Draine}(2016)}]{Hensley+Draine_2016b}
{Hensley}, B.~S., \& {Draine}, B.~T. 2016, In prep

\bibitem[{{Hoang} \& {Lazarian}(2016{\natexlab{a}})}]{Hoang+Lazarian_2016b}
{Hoang}, T., \& {Lazarian}, A. 2016{\natexlab{a}}, ArXiv e-prints,
  arXiv:1605.02828

\bibitem[{{Hoang} \& {Lazarian}(2016{\natexlab{b}})}]{Hoang+Lazarian_2016}
---. 2016{\natexlab{b}}, \apj, 821, 91

\bibitem[{{Jenkins}(2009)}]{Jenkins_2009}
{Jenkins}, E.~B. 2009, \apj, 700, 1299

\bibitem[{{Jiang} \& {Carter}(2003)}]{Jiang+Carter_2003}
{Jiang}, D.~E., \& {Carter}, E.~A. 2003, Surface Science, 547, 85

\bibitem[{{Jones}(1990)}]{Jones_1990}
{Jones}, A.~P. 1990, \mnras, 245, 331

\bibitem[{{Jones} \& {Spitzer}(1967)}]{Jones+Spitzer_1967}
{Jones}, R.~V., \& {Spitzer}, Jr., L. 1967, \apj, 147, 943

\bibitem[{{Kasama} {et~al.}(1983){Kasama}, {McLean}, {Miller}, Z., \&
  {Ward}}]{Kasama+etal_1983}
{Kasama}, A., {McLean}, A., {Miller}, W.~A., Z., M., \& {Ward}, M.~J. 1983,
  Canadian Metallurgical Quarterly, 22, 9

\bibitem[{{Keller} \& {McKay}(1997)}]{Keller+McKay_1997}
{Keller}, L.~P., \& {McKay}, D.~S. 1997, \gca, 61, 2331

\bibitem[{{Kimura}(2016)}]{Kimura_2016}
{Kimura}, H. 2016, \mnras, 459, 2751

\bibitem[{{K{\"o}hler} {et~al.}(2014){K{\"o}hler}, {Jones}, \&
  {Ysard}}]{Kohler+Ysard+Jones_2014}
{K{\"o}hler}, M., {Jones}, A., \& {Ysard}, N. 2014, \aap, 565, L9

\bibitem[{{Lodders} {et~al.}(2009){Lodders}, {Palme}, \&
  {Gail}}]{Lodders+etal_2009}
{Lodders}, K., {Palme}, H., \& {Gail}, H.-P. 2009, Landolt B{\"o}rnstein, 44

\bibitem[{{Mathis}(2000)}]{Mathis_2000}
{Mathis}, J.~S. 2000, \apj, 544, 347

\bibitem[{{Mathis} {et~al.}(1983){Mathis}, {Mezger}, \&
  {Panagia}}]{Mathis+Mezger+Panagia_1983}
{Mathis}, J.~S., {Mezger}, P.~G., \& {Panagia}, N. 1983, \aap, 128, 212

\bibitem[{{Paerels} {et~al.}(2001){Paerels}, {Brinkman}, {van der Meer},
  {Kaastra}, {Kuulkers}, {den Boggende}, {Predehl}, {Drake}, {Kahn}, {Savin},
  \& {McLaughlin}}]{Paerels+etal_2001}
{Paerels}, F., {Brinkman}, A.~C., {van der Meer}, R.~L.~J., {et~al.} 2001,
  \apj, 546, 338

\bibitem[{{Parks} {et~al.}(1990){Parks}, {Klots}, \&
  {Riley}}]{Parks+Klots+Riley_1990}
{Parks}, E.~K., {Klots}, T.~D., \& {Riley}, S.~J. 1990, \jcp, 92, 3813

\bibitem[{{Planck Collaboration} {et~al.}(2015){Planck Collaboration}, {Ade},
  {Alves}, {Aniano}, {Armitage-Caplan}, {Arnaud}, {Atrio-Barandela}, {Aumont},
  {Baccigalupi}, {Banday}, {Barreiro}, {Battaner}, {Benabed},
  {Benoit-L{\'e}vy}, {Bernard}, {Bersanelli}, {Bielewicz}, {Bock}, {Bond},
  {Borrill}, {Bouchet}, {Boulanger}, {Burigana}, {Cardoso}, {Catalano},
  {Chamballu}, {Chiang}, {Colombo}, {Combet}, {Couchot}, {Coulais}, {Crill},
  {Curto}, {Cuttaia}, {Danese}, {Davies}, {Davis}, {de Bernardis}, {de Zotti},
  {Delabrouille}, {D{\'e}sert}, {Dickinson}, {Diego}, {Donzelli}, {Dor{\'e}},
  {Douspis}, {Dunkley}, {Dupac}, {En{\ss}lin}, {Eriksen}, {Falgarone},
  {Finelli}, {Forni}, {Frailis}, {Fraisse}, {Franceschi}, {Galeotta}, {Ganga},
  {Ghosh}, {Giard}, {Gonz{\'a}lez-Nuevo}, {G{\'o}rski}, {Gregorio}, {Gruppuso},
  {Guillet}, {Hansen}, {Harrison}, {Helou}, {Hern{\'a}ndez-Monteagudo},
  {Hildebrandt}, {Hivon}, {Hobson}, {Holmes}, {Hornstrup}, {Jaffe}, {Jaffe},
  {Jones}, {Keih{\"a}nen}, {Keskitalo}, {Kisner}, {Kneissl}, {Knoche}, {Kunz},
  {Kurki-Suonio}, {Lagache}, {Lamarre}, {Lasenby}, {Lawrence}, {Leahy},
  {Leonardi}, {Levrier}, {Liguori}, {Lilje}, {Linden-V{\o}rnle},
  {L{\'o}pez-Caniego}, {Lubin}, {Mac{\'{\i}}as-P{\'e}rez}, {Maffei},
  {Magalh{\~a}es}, {Maino}, {Mandolesi}, {Maris}, {Marshall}, {Martin},
  {Mart{\'{\i}}nez-Gonz{\'a}lez}, {Masi}, {Matarrese}, {Mazzotta},
  {Melchiorri}, {Mendes}, {Mennella}, {Migliaccio}, {Miville-Desch{\^e}nes},
  {Moneti}, {Montier}, {Morgante}, {Mortlock}, {Munshi}, {Murphy}, {Naselsky},
  {Nati}, {Natoli}, {Netterfield}, {Noviello}, {Novikov}, {Novikov},
  {Oppermann}, {Oxborrow}, {Pagano}, {Pajot}, {Paoletti}, {Pasian},
  {Perdereau}, {Perotto}, {Perrotta}, {Piacentini}, {Pietrobon},
  {Plaszczynski}, {Pointecouteau}, {Polenta}, {Popa}, {Pratt}, {Rachen},
  {Reach}, {Reinecke}, {Remazeilles}, {Renault}, {Ricciardi}, {Riller},
  {Ristorcelli}, {Rocha}, {Rosset}, {Roudier}, {Rubi{\~n}o-Mart{\'{\i}}n},
  {Rusholme}, {Salerno}, {Sandri}, {Savini}, {Scott}, {Spencer}, {Stolyarov},
  {Stompor}, {Sudiwala}, {Sutton}, {Suur-Uski}, {Sygnet}, {Tauber}, {Terenzi},
  {Toffolatti}, {Tomasi}, {Tristram}, {Tucci}, {Valenziano}, {Valiviita}, {Van
  Tent}, {Vielva}, {Villa}, {Wandelt}, {Zacchei}, \& {Zonca}}]{Planck_Int_XXII}
{Planck Collaboration}, {Ade}, P.~A.~R., {Alves}, M.~I.~R., {et~al.} 2015,
  \aap, 576, A107

\bibitem[{{Poteet} {et~al.}(2015){Poteet}, {Whittet}, \&
  {Draine}}]{Poteet+Whittet+Draine_2015}
{Poteet}, C.~A., {Whittet}, D.~C.~B., \& {Draine}, B.~T. 2015, \apj, 801, 110

\bibitem[{{Quemerais} {et~al.}(1985){Quemerais}, {Seignac}, {Priol}, \&
  {Lefevre}}]{Quemerais+etal_1985}
{Quemerais}, A., {Seignac}, A., {Priol}, M., \& {Lefevre}, J. 1985, in
  Astrophysics and Space Science Library, Vol. 119, IAU Colloq. 85: Properties
  and Interactions of Interplanetary Dust, ed. R.~H. {Giese} \& P.~{Lamy},
  329--333

\bibitem[{{Robinson} \& {Holbrook}(1972)}]{Robinson+Holbrook_1972}
{Robinson}, P.~J., \& {Holbrook}, K.~A. 1972, {Unimolecular Reactions} (New
  York: Wiley)

\bibitem[{{Schalen}(1965)}]{Schalen_1965}
{Schalen}, C. 1965, \pasp, 77, 409

\bibitem[{Seah \& Dench(1979)}]{Seah+Dench_1979}
Seah, M.~P., \& Dench, W.~A. 1979, Surface and Interface Analysis, 1, 2

\bibitem[{{Tyson} \& {Miller}(1977)}]{Tyson+Miller_1977}
{Tyson}, W.~R., \& {Miller}, W.~A. 1977, Surface Science, 62, 267

\bibitem[{{Valencic} \& {Smith}(2013)}]{Valencic+Smith_2013}
{Valencic}, L.~A., \& {Smith}, R.~K. 2013, \apj, 770, 22

\bibitem[{{van Hoof} {et~al.}(2004){van Hoof}, {Weingartner}, {Martin}, {Volk},
  \& {Ferland}}]{vanHoof+etal_2004}
{van Hoof}, P.~A.~M., {Weingartner}, J.~C., {Martin}, P.~G., {Volk}, K., \&
  {Ferland}, G.~J. 2004, \mnras, 350, 1330

\bibitem[{{Wang} {et~al.}(1995){Wang}, {Cheng}, \& {Fan}}]{Wang+Cheng+Fan_1995}
{Wang}, L.-S., {Cheng}, H.-S., \& {Fan}, J. 1995, \jcp, 102, 9480

\bibitem[{{Wang} {et~al.}(2000){Wang}, {Li}, \& {Zhang}}]{Wang+Li+Zhang_2000}
{Wang}, L.-S., {Li}, X., \& {Zhang}, H.-F. 2000, Chemical Physics, 262, 53

\bibitem[{{Weingartner} \& {Draine}(2001)}]{Weingartner+Draine_2001b}
{Weingartner}, J.~C., \& {Draine}, B.~T. 2001, \apjs, 134, 263

\bibitem[{{Westphal} {et~al.}(2014){Westphal}, {Stroud}, {Bechtel}, {Brenker},
  {Butterworth}, {Flynn}, {Frank}, {Gainsforth}, {Hillier}, {Postberg},
  {Simionovici}, {Sterken}, {Nittler}, {Allen}, {Anderson}, {Ansari}, {Bajt},
  {Bastien}, {Bassim}, {Bridges}, {Brownlee}, {Burchell}, {Burghammer},
  {Changela}, {Cloetens}, {Davis}, {Doll}, {Floss}, {Gr{\"u}n}, {Heck},
  {Hoppe}, {Hudson}, {Huth}, {Kearsley}, {King}, {Lai}, {Leitner}, {Lemelle},
  {Leonard}, {Leroux}, {Lettieri}, {Marchant}, {Ogliore}, {Ong}, {Price},
  {Sandford}, {Tresseras}, {Schmitz}, {Schoonjans}, {Schreiber}, {Silversmit},
  {Sol{\'e}}, {Srama}, {Stadermann}, {Stephan}, {Stodolna}, {Sutton},
  {Trieloff}, {Tsou}, {Tyliszczak}, {Vekemans}, {Vincze}, {Von Korff},
  {Wordsworth}, {Zevin}, {Zolensky}, \& {aff14}}]{Westphal+etal_2014}
{Westphal}, A.~J., {Stroud}, R.~M., {Bechtel}, H.~A., {et~al.} 2014, Science,
  345, 786

\bibitem[{{Yang} \& {Knickelbein}(1990)}]{Yang+Knickelbein_1990}
{Yang}, S., \& {Knickelbein}, M.~B. 1990, \jcp, 93, 1533

\end{thebibliography}

\end{document}